\documentclass[aps,prb,a4paper,superscriptaddress,twocolumn,showpacs,amsmath,amssymb]{revtex4} 
\usepackage{graphicx,epsfig}
\usepackage[usenames]{color}
\usepackage{dsfont,bbm,natbib}
\usepackage{amsfonts}

\allowdisplaybreaks

\begin{document}

\newcommand{\E}{\mathcal{E}}
\newcommand{\F}{\mathcal{F}}
\newcommand{\V}{\mathcal{V}}
\newcommand{\C}{\mathcal{C}}
\newcommand{\R}{\mathcal{R}}
\newcommand{\s}{\sigma}
\newcommand{\up}{\uparrow}
\newcommand{\dw}{\downarrow}
\newcommand{\h}{\mathcal{H}}
\newcommand{\g}{\mathcal{G}^{-1}_0}
\newcommand{\D}{\mathcal{D}}
\newcommand{\A}{\mathcal{A}}
\newcommand{\K}{\textbf{k}}
\newcommand{\Q}{\textbf{q}}
\newcommand{\T}{\tau_{\ast}}
\newcommand{\io}{i\omega_n}
\newcommand{\eps}{\varepsilon}
\newcommand{\+}{\dag}
\newcommand{\su}{\uparrow}
\newcommand{\giu}{\downarrow}
\newcommand{\0}[1]{\textbf{#1}}
\newcommand{\ca}{c^{\phantom{\dagger}}}
\newcommand{\cc}{c^\dagger}
\newcommand{\da}{d^{\phantom{\dagger}}}
\newcommand{\dc}{d^\dagger}
\newcommand{\fa}{f^{\phantom{\dagger}}}
\newcommand{\fc}{f^\dagger}
\newcommand{\ha}{h^{\phantom{\dagger}}}
\newcommand{\hc}{h^\dagger}
\newcommand{\be}{\begin{equation}}
\newcommand{\ee}{\end{equation}}
\newcommand{\bea}{\begin{eqnarray}}
\newcommand{\eea}{\end{eqnarray}}
\newcommand{\ba}{\begin{eqnarray*}}
\newcommand{\ea}{\end{eqnarray*}}
\newcommand{\dagga}{{\phantom{\dagger}}}
\newcommand{\bR}{\mathbf{R}}
\newcommand{\bQ}{\mathbf{Q}}
\newcommand{\bq}{\mathbf{q}}
\newcommand{\bqp}{\mathbf{q'}}
\newcommand{\bk}{\mathbf{k}}
\newcommand{\bh}{\mathbf{h}}
\newcommand{\bkp}{\mathbf{k'}}
\newcommand{\bp}{\mathbf{p}}
\newcommand{\bL}{\mathbf{L}}
\newcommand{\bRp}{\mathbf{R'}}
\newcommand{\bx}{\mathbf{x}}
\newcommand{\by}{\mathbf{y}}
\newcommand{\bz}{\mathbf{z}}
\newcommand{\br}{\mathbf{r}}
\newcommand{\Ima}{{\Im m}}
\newcommand{\Rea}{{\Re e}}
\newcommand{\Pj}[2]{|#1\rangle\langle #2|}
\newcommand{\ket}[1]{\vert#1\rangle}
\newcommand{\bra}[1]{\langle#1\vert}
\newcommand{\setof}[1]{\left\{#1\right\}}
\newcommand{\fract}[2]{\frac{\displaystyle #1}{\displaystyle #2}}
\newcommand{\Av}[2]{\langle #1|\,#2\,|#1\rangle}
\newcommand{\Mel}[3]{\langle #1|#2\,|#3\rangle}
\newcommand{\Avs}[1]{\langle \,#1\,\rangle_0}
\newcommand{\eqn}[1]{(\ref{#1})}
\newcommand{\Tr}{\mathrm{Tr}}
\title{Efficient implementation of the Gutzwiller variational method}
\author{Nicola Lanat\`a}
\affiliation{University of Gothenburg, SE-412 96 Gothenburg, Sweden} 
\author{Hugo U. R. Strand}
\affiliation{University of Gothenburg, SE-412 96 Gothenburg, Sweden} 
\author{Xi Dai}
\affiliation{Beijing National Laboratory for Condensed Matter Physics and Institute of Physics, Chinese Academy of Sciences, Beijing 100190, China} 
\author{Bo Hellsing}
\affiliation{University of Gothenburg, SE-412 96 Gothenburg, Sweden} 
\date{\today} 
\begin{abstract} 
We present a self-consistent numerical approach to solve the Gutzwiller
variational problem for general multi-band models with arbitrary on-site interaction.
The proposed method generalizes and improves the procedure derived by Deng 
\emph{et al.}, Phys.~Rev.~B.~{\bf 79}~075114~(2009),
overcoming the restriction to density-density interaction without increasing the 
complexity of the computational algorithm. 
Our approach drastically reduces the problem  
of the high-dimensional Gutzwiller minimization by mapping it to a 
minimization only in the variational density matrix, 
in the spirit of the Levy and Lieb formulation of DFT.
For fixed density the Gutzwiller renormalization matrix is determined 
as a fixpoint of a proper functional, whose evaluation only requires 
ground-state calculations of matrices defined in the Gutzwiller variational space.
Furthermore, the proposed method is able to account for the
symmetries of the variational function in a controlled way, 
reducing the number of variational parameters.
After a detailed description of the method we present calculations for
multi-band Hubbard models with full (rotationally invariant)
Hund's rule on-site interaction.
Our analysis shows that the numerical algorithm is very efficient, stable
and easy to implement.
For these reasons this method is particularly suitable for first principle studies 
-- e.g., in combination with DFT -- of many complex real materials,  
where the full intra-atomic interaction is important to obtain correct results.

\end{abstract}

\maketitle

\section{Introduction}

In the 60th's Martin Gutzwiller published a series of papers~\cite{Gutzwiller1,Gutzwiller2,Gutzwiller3}
where he introduced a variational method for studying ferromagnetism in transition metals.
His brilliant idea was to variationally determine a projected wavefunction represented as
\be
|\Psi_G\rangle = \prod_{\bR}\, \mathcal{P}_\bR\,|\Psi_0\rangle\,, \label{GWFch1}
\ee
where the local operators $\mathcal{P}_\bR$ improve the non-interacting wavefunction $|\Psi_0\rangle$
in accordance with the on-site interaction by modifying the weight of local electronic configurations.

In spite of its simplicity, the average values of any operator on $|\Psi_G\rangle$ 
can only be computed numerically for realistic lattice models, e.g., using
variational Monte Carlo.~\cite{Yokuyama,Manuela} 
For this reason, Gutzwiller introduced an approximate scheme, known as the 
Gutzwiller approximation, to compute these average values analytically.  
Successively, the development of dynamical mean field theory (DMFT)~\cite{DMFT}
has brought additional insights into the physical 
meaning of the Gutzwiller approximation. 
In fact, Metzner and Vollhardt showed that this approximation is exact in the limit of infinite 
coordination lattices,~\cite{Metzner-Vollhardt-PRL,Metzner-Vollhardt-PRB,Gebhard1} 
where the single-particle self-energy becomes purely local in space.~\cite{Muller}

Since its introduction the Gutzwiller wavefunction and approximation 
have proven to be very important tools to study strongly correlated systems.
The understanding of many basic concepts, 
such as the Brinkman-Rice scenario for the Mott transition,~\cite{brinkman&rice} 
came originally from calculations based on the Gutzwiller method.
From the computational point of view the list of 
interesting results that have been obtained 
by means of the Gutzwiller approximation~\cite{Gutzwiller1,Gutzwiller2,Gutzwiller3} 
and its respective 
generalizations~\cite{Dzierzawa,Gebhard1,metr,PhysRevLett.105.076401,td-gutz-transport}
is impressively long, hence impossible to cite in an exhaustive way.
Furthermore, the Gutzwiller approximation can be naturally combined with  
density functional theory (DFT),~\cite{HohenbergandKohn,KohnandSham} applying, e.g.,
the local density approximation (LDA)~\cite{LDA} for the exchange and correlation. 
LDA+Gutzwiller (LDA+G)~\cite{poorchguy,Fang} 
has proven to be a powerful scheme
for the study of real strongly-correlated metallic materials;~\cite{Fang} 
giving a more accurate description than LDA+U,~\cite{LDA+U}
comparable to LDA+DMFT~\cite{LDA+U+DMFT} for ground state properties.
Finally, the range of application of the Gutzwiller method has recently been
extended to out-of-equilibrium calculations, 
such as electron transport across quantum dot systems~\cite{metr} and quench dynamics in
correlated electron systems.~\cite{PhysRevLett.105.076401}

The Gutzwiller variational 
method requires a number of preliminary technical steps 
in order to make it really flexible and able to cope with systems of interest, 
as the number of variational parameters scales exponentially with the number of correlated 
orbitals involved in the calculation.
This scaling is already problematic for transition metal systems with correlated $d$ orbitals
if the required minimization is performed in a naive way.
Based on the formalism introduced by B\"unemann,~\cite{Gebhard}
Deng \emph{et al.}~\cite{Fang} recently derived 
a self-consistent numerical method that allows to efficiently perform calculations
even for $d$-orbital systems.
The only limitation of this approach is the restriction to density-density type of 
interactions, which is actually due to the employed formalism and does not stem 
from the numerical approach in itself.

However, in order to properly describe the physics of several strongly correlated materials, 
the full intra-atomic interaction is needed; not only
its density-density component.
The rotational invariant 
on-site Coulomb and exchange interaction is generally modeled in terms 
of the so called Kanamori parameters~\cite{Kanamori}
commonly referred to with the symbols $U$ and $J$. 
The strength of the two-electron spin-exchange interaction is determined
by the parameter $J$.
In transition metal oxides with partly filled $d$-shells the off-diagonal interactions -- 
exchange coupling, spin-flip and pair hopping -- are crucial.
For example, 
the metallic property of SrVO$_{3}$ can not be reproduced from theory without 
accounting for spin-exchange interaction.~\cite{Werner-PRB2009}
Several theoretical model studies points in the same direction. 
To mention a few, the transition from the paramagnetic to the ferromagnetic phase 
for multiband systems requires finite $J$,~\cite{Gebhard}
the spin-freezing transition predicted for multiband systems, 
which is expected to influence the Mott transition,~\cite{Costi-PRL2007}
takes place when $0<J/U<1/3$ and is absent for $J=0$.~\cite{Werner2008aa}
Previous Gutzwiller model studies~\cite{Gebhard} and the present work
show that the critical $U$ required for the Mott insulator transition is substantially 
reduced when increasing the ratio $J/U$ from zero.

Motivated by the above examples we believe that an implementation of the 
Gutzwiller variational method valid for general on-site interactions 
and, at the same time, numerically efficient 
would constitute an important progress.
Formal advancements pointing in this direction have lately been conceived by
several authors,~\cite{PhysRevB.55.4011,PhysRevB.47.8594,Gebhard,Attaccalite,wang:2006,fab,lanata,mybil} although these progresses have been hampered 
by the lack of efficient numerical algorithms applicable to the most general case.
In particular, Fabrizio and collaborators~\cite{Attaccalite,ferreroPRB,fab,lanata,mybil,mythesis} derived a 
mathematical formulation of the problem whose complexity is unaffected by the 
form of the on-site interactions and furthermore allows to easily incorporate
symmetries into the variational function from the onset.

The main goal of this work consists of merging together the general formalism developed 
by Fabrizio and collaborators~\cite{Attaccalite,ferreroPRB,fab,lanata,mybil} mentioned above and 
the numerical procedure derived by Deng \emph{et al.},~\cite{Fang}
overcoming the restriction to density-density interaction without increasing the 
complexity of the computational algorithm. 
Furthermore, the fully rotational invariant on-site interaction enables us to construct 
the variational wavefunction using the orbital rotational symmetry, which is instead broken
if the off-diagonal terms are neglected.
It will be shown that
this extra symmetry can be used to reduce the number of variational parameters.

The outline of this paper is as follows.
In Sec.~\ref{method} the Gutzwiller problem for a general tight binding Hamiltonian
is introduced.
In Sec.~\ref{phiform} the employed formulation of the 
Gutzwiller method~\cite{Attaccalite,fab,lanata,mybil,mythesis}
is summarized. In particular, in Sec.~\ref{dprojcase} it is shown that this formulation
provides a natural extension of the formalism of Ref.~\onlinecite{Fang}, having  
the same mathematical structure in the special case of density-density on-site interactions.
In Sec.~\ref{symphymat} we discuss the implementation of symmetries of 
the wavefunctions.
In Sec.~\ref{enopt} the numerical procedure to 
minimize the Gutzwiller energy is described in detail.
In Sec.~\ref{ldapg} we briefly discuss how the proposed method can be adapted to a
of LDA+G type of calculation.
In Sec.~\ref{modres} we prove the reliability of the method presenting 
a comparison with other methods for the cases of two and five orbitals. 
Furthermore we discuss several technical
details of the numerical procedure, such as convergence properties and
computational speed.
Finally, Sec.~\ref{concl} is devoted to the conclusions.

\section{The Gutzwiller method}\label{method}

Let us consider the general tight binding Hamiltonian 
\bea
\hat{\h}&=&\sum_{\bR\neq\bRp}\sum_{\alpha\beta}\,t_{\bR\bRp}^{\alpha\beta}\,
\cc_{\bR\alpha}\ca_{\bRp\beta}\nonumber\\
&+&
\label{hamgen}
\sum_{\bR}\sum_{\Gamma\Gamma'}U(\bR)_{\Gamma\Gamma'}\,\ket{\Gamma,\bR}
\bra{\Gamma',\bR}
\\&\equiv&
\hat{T}+\hat{{H}}_{\text{loc}}\nonumber
\,,
\eea
where $\cc_{\bR\alpha}$ creates an electron 
in state $\alpha$ (where $\alpha$ labels both the spin $\sigma$ and 
the orbital $a$ at site $\bR$)
and $|\Gamma,\bR\rangle$ are many-body Fock states expressed in the 
$c_{\bR\alpha}$-basis. 
These states are defined by the occupation numbers 
$n_{\alpha}(\Gamma,\bR)\in\{0,1\}$ 
where $\alpha$ runs over integer numbers from 1 to $M$, 
$M$ being the number of on-site single particle states, 
\be
|\Gamma,\bR\rangle = \left(c^\dagger_{\bR 1}\right)^{n_{1}(\Gamma,\bR)}\!\!\!\!\!\!.\,.\,.\;
\left(c^\dagger_{\bR M}\right)^{n_{M}(\Gamma,\bR)}\,|0\rangle\,.
\label{gammastates}
\ee
Thus the number of Fock states is $2^{M}$. The Hermitian matrix $U(\bR)$ 
represents the local terms, interaction and crystal fields, 
in the $\cc_{\bR\alpha}$-basis, i.e., 
the same basis in which $\hat{T}$ was defined in Eq.~\eqref{hamgen}.
This basis will henceforth be denoted as the \emph{original} basis.

The structure of the Gutzwiller variational function is given by Eq.~\eqref{GWFch1},
where $\vert\Psi_0\rangle$ is an uncorrelated variational wavefunction,
that satisfies Wick's theorem, and 
$\mathcal{P}_\bR$ is a general operator acting on the local configurations 
at site $\bR$
\be
\mathcal{P}_\bR =
\sum_{\Gamma\Gamma'}\lambda(\bR)_{\Gamma\Gamma'}\,\ket{\Gamma,\bR}
\bra{\Gamma',\bR}\,,
\label{PRch1}
\ee
where the $2^{M}\!\times\! 2^{M}$ matrix $\lambda(\bR)$, assumed to be real in this work, 
contains all the variational parameters needed to define the operator $\mathcal{P}_\bR$.

In general, average values of operators 
with respect to $|\Psi_G\rangle$ must be computed numerically unless the lattice
has infinite coordination number,
in which case they can be evaluated analytically if
the following equations -- commonly named Gutzwiller constraints -- 
are satisfied:
\bea
\Av{\Psi_0}{\mathcal{P}^\dagger_\bR\,\mathcal{P}^\dagga_\bR} &=& 1\label{construnoch1}\\
\Av{\Psi_0}{\mathcal{P}^\dagger_\bR\,\mathcal{P}^\dagga_\bR\,\mathcal{C}_{\bR}} &=& 
\Av{\Psi_0}{\mathcal{C}_{\bR}}\,,\label{constrduech1}
\eea 
where $\mathcal{C}_{\bR}$ is the local single-particle density-matrix operator 
with elements 
$\cc_{\bR\alpha}\ca_{\bR\beta}$.

The variational problem to solve amounts to variationally determine
both $\vert\Psi_0\rangle$ and $\mathcal{P}_\bR$ by minimizing the average 
value of the Hamiltonian~[Eq.\eqref{hamgen}]
\be
\mathcal{E}_{\text{var}}\left[\mathcal{P},\Psi_0\right]=
\Av{\Psi_0}{\mathcal{P}^\dagger\hat{\h}\mathcal{P}^\dagga\!\!}
\label{varen}
\ee
fulfilling Eqs.~(\ref{construnoch1}-\ref{constrduech1}), where we have introduced
\be
\mathcal{P}\equiv\prod_{\bR}\,\mathcal{P}_\bR.
\ee
For a general tight binding model~[Eq.~\eqref{hamgen}] this problem is 
complicated for two reasons: 
(i) $|\Psi_0\rangle$ and $\mathcal{P}$ are not independent variables because of 
the Gutzwiller constraints~[Eqs.~(\ref{construnoch1}-\ref{constrduech1})], 
(ii) the number of variational parameters scales exponentially 
with the number of orbitals.

\section{Reformulation of the Gutzwiller problem}\label{phiform}

In this section we briefly summarize the reformulation of the Gutzwiller 
problem derived in Refs.~\onlinecite{lanata},~\onlinecite{mybil},
and we show its formal analogy with
the formulation of B\"unemann and Weber~\cite{Gebhard} 
in the special case of pure density-density local interaction.

\subsection{The mixed-basis representation}\label{mixbassect1}

Let us introduce the so-called \emph{natural-basis}~\cite{fab} 
operators $\da_{\bR\alpha}$, i.e., the operators such that  
\be
\Av{\Psi_0}{d^\dagger_{\bR\alpha}d^\dagga_{\bR\beta}} = \delta_{\alpha\beta}\,n^0_{\bR\alpha} 
\equiv n^0_{\alpha\beta}(\bR)
\quad\forall\,\alpha,\beta
\label{C-avch1}
\ee
where $0\leq n^0_{\bR\alpha}\leq 1$ are the eigenvalues of 
the local density matrix
\be
\Av{\Psi_0}{c^\dagger_{\bR\alpha}c^\dagga_{\bR\beta}}
\equiv\bar{\rho}^0_{\alpha\beta}(\bR)\,.
\label{ldmcbasis}
\ee
Notice that the natural-basis operators are always well defined as 
$\bar{\rho}^0(\bR)$ is Hermitian, implying that there always exists a 
unitary transformation $\mathcal{U}^\bR$ such that
\be
d^\dagger_{\bR\alpha}=\sum_\beta\mathcal{U}^\bR_{\beta\alpha}\,c^\dagger_{\bR\beta}\,.
\label{orinnat}
\ee

Instead of expressing the  Gutzwiller projector in terms of the original basis
as in Eq.~\eqref{PRch1} we adopt the following 
mixed \emph{original-natural}~\cite{lanata} basis form
\be
\mathcal{P}_\bR = 
\sum_{\Gamma n}\lambda(\bR)_{\Gamma n}\,\ket{\Gamma,\bR}
\bra{n,\bR}
\label{mixbaproj}
\ee
where, by assumption, $|\Gamma,\bR\rangle$ are Fock states in the 
original $c_{\bR\alpha}$-basis ,  
while $|n,\bR\rangle$ are Fock states in the natural basis, namely in terms of 
the $d_{\bR\alpha}$-operators. In other words a generic state $|n,\bR\rangle$ 
is identified by the occupation numbers 
$n_{\beta}(n,\bR)\in\{0,1\}$ -- with $\beta\in\{1,..,M\}$ --
and has the explicit expression 
\be
|n,\bR\rangle = \left(d^\dagger_{\bR 1}\right)^{n_1(n,\bR)}\!\!\!\!\!\!.\,.\,.\;
\left(d^\dagger_{\bR M}\right)^{n_M(n,\bR)}\,|0\rangle\,.
\ee
For later convenience 
we adopt the convention that the order of the $|\Gamma,\bR\rangle$ and the 
$|n,\bR\rangle$ states is the same.
For instance, if the second $\Gamma$-vector 
in Eq.~\eqref{mixbaproj} is  
$c^\dagger_{1\uparrow}c^\dagger_{2\downarrow}\,|0\rangle$ then
the second $n$-vector is $d^\dagger_{1\uparrow}d^\dagger_{2\downarrow}\,|0\rangle$.

\subsection{The $\phi$-matrix}\label{mixbassect2}

Let us introduce the uncorrelated occupation-probability matrix $P^0(\bR)$~\cite{fab}
with elements  
\bea
[P^0(\bR)]_{n n'} &\equiv& \Av{\Psi_0}{|n',\bR\rangle\langle n,\bR|}\nonumber\\
&=& \delta_{n n'}\,P^0_{n}(\bR)\,,\label{P0ch1}
\eea
where 
\be
P^0_{n}(\bR) = \prod_{\beta=1}^{M}\, \left(n^0_{\bR\beta}\right)^{n_{\beta}(n,\bR)}\!
\left(1-n^0_{\bR\beta}\right)^{1-n_{\beta}(n,\bR)}\,.\label{Pn0ch1}
\ee
We remind that $n^0_{\bR\beta}$ are the elements of the diagonal density matrix
of Eq.~\eqref{C-avch1},
and denote the occupation numbers of the natural states $\beta$.
We also introduce the matrix representation of the operators $d^\dagga_{\bR\beta}$
and $c^\dagga_{\bR\beta}$
\bea
d^\dagga_{\bR\beta}\rightarrow \left(d^\dagga_{\bR\beta}\right)_{n n'} 
&=& \langle n,\bR|d^\dagga_{\bR\beta}|n',\bR\rangle\\
c^\dagga_{\bR\beta}\rightarrow \left(c^\dagga_{\bR\beta}\right)_{\Gamma\Gamma'} 
&=& \langle \Gamma,\bR|c^\dagga_{\bR\beta}|\Gamma',\bR\rangle\,.
\eea
Notice that, if we respect the convention that the order of the 
$|\Gamma,\bR\rangle$ and the 
$|n,\bR\rangle$ states is the same, we have that
\be
\left(c^\dagga_{\bR\beta}\right)_{ij}\!\!=
\left(d^\dagga_{\bR\beta}\right)_{ij}\equiv \left(f^\dagga_\beta\right)_{ij}
\quad\forall\,\beta,\,i,j\,.
\label{fmats}
\ee

We now define the matrices $\lambda(\bR)$ and $U(\bR)$ with elements 
$\lambda_{\Gamma n}(\bR)$~[Eq.~\eqref{mixbaproj}]
and $U_{\Gamma\Gamma'}(\bR)$~[Eq.~\eqref{hamgen}].
Notice that $\lambda(\bR)$ is defined in the original-natural and $U(\bR)$
is defined in the original-original basis.
With the above definitions, 
the expectation value of any local observable can be calculated as 
\be
\Av{\Psi_0}{\mathcal{P}^\dagger\hat{\mathcal{O}}(\bR)\mathcal{P}^\dagga\!\!}
=\Tr\left(P^0(\bR)\lambda^\dagger\!(\bR)\,\mathcal{O}(\bR)\,\lambda(\bR)\right)\,,
\label{avlocold}
\ee
where
\be
\mathcal{O}_{\Gamma\Gamma'}(\bR)=\langle\Gamma|\hat{\mathcal{O}}|\Gamma'\rangle\,,
\ee
and the Gutzwiller constraints~[Eqs.~(\ref{construnoch1}-\ref{constrduech1})] 
can be written as
\bea
\Tr\!\left(P^0(\bR)\lambda^\dagger\!(\bR)\lambda(\bR)\right) \!\!\!\!&=&\!\!\! 1\label{uno-appch1}\\
\Tr\!\left(P^0(\bR)\lambda^\dagger\!(\bR)\lambda(\bR)
f^\dagger_{\alpha}f^\dagga_{\beta}\right)
\!\!\!\!&=&\!\!\! \Av{\Psi_0}{\!d^\dagger_{\bR\alpha}\!d^\dagga_{\bR\beta}\!}.
\eea

The formalism is further simplified by defining the matrix
\be
\phi(\bR) = \lambda(\bR)\,\sqrt{P^0(\bR)}\label{phich1}\,,
\ee
that was introduced in Ref.~\onlinecite{lanata}.
The expectation value of any local observable is given by
\be
\Av{\Psi_0}{\mathcal{P}^\dagger\hat{\mathcal{O}}(\bR)\mathcal{P}^\dagga\!\!}
=\Tr\left(\phi(\bR)^\dagger\,\mathcal{O}(\bR)\,\phi(\bR)\right)\,.
\label{avloc}
\ee
The Gutzwiller constraints take the form
\bea
\Tr\left(\phi^\dagger(\bR)\phi(\bR)\right) \!\!&=&\!\! 1\,,\label{uno-app-bisch1}\\
\Tr\left(\phi^\dagger(\bR)\phi(\bR)\,f^\dagger_{\alpha}f^\dagga_{\beta}\right) 
\!\!&=&\!\!\Av{\Psi_0}{d^\dagger_{\bR\alpha}d^\dagga_{\bR\beta}}\nonumber\\
\!\!&\equiv&\!\!\delta_{\alpha\beta}\,n^0_{\bR\alpha}\,,
\label{due-app-bis1ch1}
\eea
and the variational energy~[Eq.~\eqref{varen}] is, in the 
Gutzwiller approximation, given by~\cite{lanata}
\bea
\mathcal{E}_{\text{var}}&=& \sum_{\bR\bRp}\sum_{\gamma\delta}\,\tilde{t}_{\bR\bRp}^{\gamma\delta}
\,\Av{\Psi_0}{\dc_{\bR,\gamma}\da_{\bRp,\delta}} \nonumber\\
&&\,+\, \sum_{\bR}\Tr\left(\phi(\bR)^\dagger\, U(\bR)\,\phi(\bR)\right)\,,
\label{variationalenergych1}
\eea
where
\bea
\tilde{t}_{\bR\bRp}^{\gamma\delta}&\equiv&\sum_{\alpha\beta}
t_{\bR\bRp}^{\alpha\beta}\,
\R(\bR)_{\alpha\gamma} \R(\bRp)_{\beta\delta}
\label{ttilde}\\
\R(\bR)_{\alpha\beta} &=& 
\frac{\Tr(\phi^{\dagger}(\bR)\,\fc_{\alpha}\,\phi(\bR)\,\fa_{\beta})}
{\sqrt{n^0_{\beta}(\bR)(1-n^0_{\beta}(\bR))}}
\,.\label{Z-newch1}
\eea

In conclusion,
within the formalism summarized in this section, the variational 
energy is a functional of $\phi(\bR)$ and $\ket{\Psi_0}$,
to be minimized fulfilling the Gutzwiller 
constraints~[Eqs.~(\ref{uno-app-bisch1}-\ref{due-app-bis1ch1})].

\subsection{Diagonal projector as a particular case}\label{dprojcase}

Let us assume that the coefficients of the matrix $\lambda$ that define the 
projector $\mathcal{P}$ in Eq.~\eqref{PRch1} are diagonal 
\be
\lambda(\bR)_{\Gamma\Gamma'}=\delta_{\Gamma\Gamma'}\,\lambda(\bR)_{\Gamma\Gamma}\,,
\label{ppp1}
\ee
and that the \emph{original} basis coincides with the \emph{natural} basis
\bea
\Av{\Psi_0}{c^\dagger_{\bR\alpha}c^\dagga_{\bR\beta}}&=&
\delta_{\alpha\beta}\,\Av{\Psi_0}{c^\dagger_{\bR\alpha}c^\dagga_{\bR\alpha}}
\nonumber\\
&\equiv&\Av{\Psi_0}{d^\dagger_{\bR\alpha}d^\dagga_{\bR\beta}}
\,,
\label{ppp2}
\eea
From Eqs.~(\ref{ppp1}-\ref{ppp2}) we have that 
\be
\phi(\bR)_{ij}=\delta_{ij}\,\phi(\bR)_{ii}\,,
\label{diagphi}
\ee 
and, consequently, the following equation hold for all local operators 
$\mathcal{O}(\bR)$
\be
\Tr\left(\phi^\dagger(\bR)\,\phi(\bR) \,\mathcal{O}(\bR)\right)
=\Tr\left(\phi^\dagger(\bR)\,\mathcal{O}(\bR)\,\phi(\bR)\right)\,.
\label{intparticular}
\ee
From Eq.~\eqref{intparticular} follows that the equations that characterize
the Gutzwiller problem~[Eqs.~(\ref{uno-app-bisch1}-\ref{Z-newch1})]
can be evaluated in terms of 
the variational parameters $\sqrt{m_{\Gamma}(\bR)}$ defined by 
B\"unemann in Ref.~\onlinecite{Gebhard}
\bea
\phi(\bR)_{\Gamma\Gamma}\!\!&=&\!\!\sqrt{\Av{\Psi_0}{\mathcal{P}^\dagger
|\Gamma,\bR\rangle\langle\Gamma,\bR|\mathcal{P}^\dagga\!\!}}\nonumber\\
\!\!&\equiv&\!\! \sqrt{m_{\Gamma}(\bR)}\,.\label{mparxidai}
\eea

A crucial observation in this work is that in the general case considered
here, in which Eqs.~(\ref{ppp1}-\ref{ppp2}) are not assumed, 
Eqs.~(\ref{uno-app-bisch1}-\ref{Z-newch1}) are expressed in terms of 
quadratic forms of the matrix elements of $\phi(\bR)$ instead of 
$\sqrt{m_{\Gamma}(\bR)}$, but in a formally identical way.

We conclude this section observing that the \emph{physical} density 
matrix of the system
\bea
\rho_{\alpha\beta}(\bR)&\equiv&\Av{\Psi_0}{\mathcal{P}^\dagger
c^\dagger_{\bR\alpha}c^\dagga_{\bR\beta}\mathcal{P}^\dagga\!\!}
\nonumber\\&=&
\Tr\left(\phi^\dagger(\bR)\,f^\dagger_{\alpha}f^\dagga_{\beta}\,\phi(\bR)\right)
\label{ordm}
\eea
is \emph{not} equal to the so called \emph{variational} density matrix
\bea
n^0_{\alpha\beta}(\bR)&\equiv&\Av{\Psi_0}{d^\dagger_{\bR\alpha}d^\dagga_{\bR\beta}}
\nonumber\\&=&
\Tr\left(\phi^\dagger(\bR)\,\phi(\bR)\,f^\dagger_{\alpha}f^\dagga_{\beta}\right) 
\label{vardm}
\eea
when Eqs.~(\ref{ppp1}-\ref{intparticular}) do not hold.
In the general case the distinction between variational density matrix 
and physical density matrix needs to be taken into account.

\section{Symmetries of the variational function and $\phi$ matrix}\label{symphymat}

In this section we discuss in detail how to 
build symmetries in the Gutzwiller variational function.
The site label $\bR$ is dropped for simplicity.
The procedure discussed here extends the method discussed in Ref.~\onlinecite{mybil}
to \emph{general} point symmetry groups.
The problem amounts to define the form 
of the $\phi$-matrix such that $\ket{\Psi_G}$
is invariant under the action of a matrix representation of a group $G$ in the many-body
$\bR$-local space.
The transformation law 
\bea
g\,\cc_{\alpha}\,g^{-1}
\!\!&=&\!\!\sum_\beta {D}^\dagga_{\beta\alpha}(g)\,\cc_{\beta}\quad\forall g\in G
\nonumber\\
g\,|0\rangle\!\!&=&\!\!|0\rangle
\label{trbands}
\eea
defines a representation~\cite{Wigner} $R(G)$ of $G$ in the local Hilbert space generated by 
the Fock configurations $\Gamma$, see Eq.~\eqref{gammastates}.
\be
g\,|\Gamma\rangle=\sum_{\Gamma'} R_{\Gamma'\Gamma}(g)\,|\Gamma'\rangle\,.
\ee
In the \emph{original-original} basis~[Eq.~\eqref{PRch1}] the invariance condition
\be
g\,\mathcal{P}\,g^{-1}=\mathcal{P}\quad\forall g \in G
\ee
is equivalent to
\be
[\lambda,R(g)]=0\quad\forall g \in G\,.
\label{ii1}
\ee

Let us \emph{assume} that the most general transformation $U$
that relates the original and the natural basis
\bea
\dc_\alpha\equiv U\,\cc_{\alpha}\,U^\dagger
&=&\sum_\beta \mathcal{U}_{\beta\alpha}\,\cc_{\beta}
\nonumber\\
U\,|0\rangle&=&|0\rangle
\nonumber\\
U\,|\Gamma\rangle&=&|n\rangle
\,,
\label{ornattransf}
\eea
commutes with $G$, i.e., that
\be
[U,R(g)]=[U^\dagger,R(g)]=0\quad\forall g\in G
\,.
\label{assumsym}
\ee
It can be shown, see appendix~\ref{appsym_sub}, that Eq.~\eqref{assumsym} is verified 
for every group $G$ whose elements do not mix configurations that belong to different
eigenspaces of the number operator $\hat{N}$. 
This assumption is obviously verified by every geometry group, 
but excludes, for instance, the particle-hole transformation.
A first consequence of Eq.~\eqref{assumsym} is that the matrix 
$\lambda$ has the same form 
in the original-natural and in the original-original representation.
In fact
\be
\mathcal{P}\equiv\sum_{\Gamma\Gamma'}\lambda_{\Gamma\Gamma'}\ket{\Gamma}\bra{\Gamma'}
=\sum_{\Gamma n}\left(\lambda U\right)_{\Gamma n}\ket{\Gamma}\bra{n}\,,
\ee 
and from Eqs.~\eqref{ii1} and \eqref{assumsym} we have that 
\be
[\lambda U,R(g)]=0\quad\forall g\in G\,.
\ee

Notice that from the assumed invariance of $|\Psi_0\rangle$
respect to $G$ we have that, $\forall g \in G$,
\bea
{P}_{\Gamma\Gamma'}&\equiv&\Av{\Psi_0}{|{\Gamma'}\rangle\langle{\Gamma}|}
\nonumber\\
&=&\Av{\Psi_0}{|g\,{\Gamma'}\rangle\langle g\,{\Gamma}|}
\nonumber\\
&=&\left({R}^\dagger(g)\,{P}\,{R}(g)\right)_{\Gamma\Gamma'}\,.
\label{ii2}
\eea
Using Eq.~\eqref{ornattransf} we can easily express $P$ in terms of 
$P^0$ as follows
\bea
{P}=U P^0 U^\dagger\,,
\label{ii3}
\eea
where
\be
{P}^0_{ij}\equiv\Av{\Psi_0}{|{n}_j\rangle\langle{n}_i|}\,,
\ee
so that, combining Eq.~\eqref{ii2} and Eq.~\eqref{ii3}, we obtain that 
\be
P^0=\left(U^\dagger R^\dagger(g)U\right)P^0\left(U^\dagger R(g)U\right)\,.
\label{p0invstep}
\ee
Eq.~\eqref{p0invstep} is equivalent, 
because of Eq.~\eqref{assumsym}, to the following invariance relation
for $P^0$:
\be
[P^0,R(g)]=0\quad\forall g\in G\,.
\ee

From the above considerations and Eq.~\eqref{phich1} we can conclude that
$\phi$ satisfies the same invariance relation of the 
$\lambda$ coefficients of the Gutzwiller projector in the original-original
basis, i.e., that
\be
[\phi,R(g)]=0\quad\forall g\in G\,.
\label{symphi}
\ee
In other words, we have proven that \emph{$\phi$ has the same 
form of $\lambda$} expressed in the original-original Fock 
representation~[Eq.~\eqref{PRch1}].

The set of all the ${\phi}$ matrices that satisfy
Eq.~\eqref{symphi} is a linear space $\mathcal{V_\phi}$. 
Consequently, there exists a basis of matrices $\{\phi_k\}$ such that
\bea
&&[\phi_k,{R}(g)]=0\quad\forall g\in G \label{commRg}\\
&&\phi=\sum_k c_k\,\phi_k\,. \label{phikbasis}
\eea
In this work we assume that $c_k$ and $\phi_k$ are real.  
This means that $\mathcal{V_\phi}$ is a linear space over the field
of \emph{real} numbers.
Notice that this does not restrict the variational freedom
as long as the local interaction $\hat{H}_{\text{loc}}$ is real. 
This excludes, for example, the spin-orbit coupling.

\subsection{Calculation of $\{\phi_k\}$}

In order to calculate $\{\phi_k\}$ it is convenient to 
apply a similarity transformation $V$ to $R(G)$
\be
R^V(g)=V R(g) V^\dagger\quad\forall g\in G
\label{transftobar}
\ee
with the property to decompose $R(G)$ in irreducible representations.~\cite{Wigner}
More precisely, we need to calculate a unitary matrix $V$ such that $R^V(G)$
is of the form
\be
R^V(g)=
\left(
\begin{array}{ccc}
R^V_1(g) & \cdots\ & 0 \\
\vdots  & \ddots & \vdots \\
0 & \cdots & R^V_s(g)  \\
\end{array}
\right)\quad\forall g\in G\,,
\label{formRg_V}
\ee
where (i) the representations $R^V_i(G)$ are irreducible for all $i\in\{1,..,s\}$, 
(ii) if two representations $R^V_{i},R^V_{j}$ are equivalent then they are 
also equal.
In appendix~\ref{appsym} we derive
a possible procedure to calculate explicitly the similarity transformation 
$V$ utilized in this section for a general geometry group $G$.

Let us consider the linear space $\mathcal{\bar{V}}^V_\phi$ 
(over the \emph{real} field) of the complex matrices $\bar{\phi}^V$ that satisfy 
the following equation
\be
[R^V(g),\bar{\phi}^V]=0\quad\forall g \in G\,.\label{rvphi}
\ee
It can be easily proven by means of the Schur lemma~\cite{Wigner} that 
the most general $\bar{\phi}^V\in\mathcal{\bar{V}}^V_\phi$ is of the form
\be
\bar{\phi}^V=
\left(
\begin{array}{ccc}
p^V_1 & \cdots\ & 0 \\
\vdots  & \ddots & \vdots \\
0 & \cdots & p^V_{r}  \\
\end{array}
\right)\,,
\label{formphi_V}
\ee
where the blocks $k\in\{1,..,s\}$ correspond to inequivalent representations of $G$, and
each block is of the form
\be
p^V_k = \left(
\begin{array}{ccc}
r_{11}\mathbbm{1}_{d_k} & \cdots & r_{1 n_k}\mathbbm{1}_{d_k} \\
\vdots  & \ddots & \vdots \\
r_{n_k 1}\mathbbm{1}_{d_k} & \cdots & r_{n_k n_k}\mathbbm{1}_{d_k} \\
\end{array}
\right)
\,,
\label{formphi_V_equivalent}
\ee
$\mathbbm{1}_{d_k}$ being identity matrices of size $d_k\times d_k$, $d_k$ being
the dimension of each one of the irreducible equivalent representations of $G$ repeated 
in the $k$-th block, and $r_{ij}$ being independent \emph{complex} numbers.
Eqs.~\eqref{formphi_V} and \eqref{formphi_V_equivalent} 
allow to define straightforwardly a basis $\{\bar{\phi}^V_k\}$ 
of $\mathcal{\bar{V}}^V_\phi$.

Let us define now the linear space of real matrices 
$\bar{\mathcal{V}}_\phi$ generated by the set
of matrices $\{\bar{\phi}_k\}$ obtained as
\be
\bar{\phi}_k\equiv V^\dagger \bar{\phi}^V_k V\label{vphiv}\,.
\ee
The linear space $\mathcal{V}_\phi$ that we need, see Eq.~\eqref{symphi}, 
is obtained as 
\be
\mathcal{V}_\phi=\mathcal{\bar{V}}_\phi\cap\mathcal{W}_{\mathbb{R}}\,,
\ee
where $\mathcal{W}_{\mathbb{R}}$ is the linear space of all 
\emph{real} matrices. 
It is convenient, finally, to orthonormalize the basis set $\{\phi_k\}$ of 
$\mathcal{V}_\phi$ in order to have
\be
\Tr(\phi^\dagger_i\phi_j)=\delta_{ij}\,.\label{orthphik}
\ee

\subsection{Independent variational parameters}\label{tensors}

All the relevant quantities that define the variational 
energy, i.e., (i) the Gutzwiller 
constraints~[Eqs.~(\ref{uno-app-bisch1}-\ref{due-app-bis1ch1})],
(ii) the $\R$ matrices~[Eq.~\eqref{Z-newch1}] and (iii) the local-interaction 
energy~[Eq.~\eqref{variationalenergych1}] can be expressed in terms of quadratic 
forms in the $c_k$ coefficients defined in Eq.~\eqref{phikbasis} as follows:
\bea
\R_{\alpha\beta}&=&\sum_{ij}c_ic_j
\frac{\text{Tr}\left(\phi^\dagger_i\fc_\alpha\phi^\dagga_j\fa_\beta\right)}
{\sqrt{n^0_\beta(1-n^0_\beta)}}\nonumber\\
&\equiv&\sum_{ij}c_ic_j\frac{M^{ij}_{\alpha\beta}}{\sqrt{n^0_\beta(1-n^0_\beta)}}
\nonumber\\
&=&\sum_{ij}c_ic_j
\,\frac{1}{2}
\frac{M^{ij}_{\alpha\beta}+M^{ji}_{\alpha\beta}}{\sqrt{n^0_\beta(1-n^0_\beta)}}
\nonumber\\
&\equiv&\Av{c}{ \frac{{M}^S_{\alpha\beta}}{\sqrt{n^0_\beta(1-n^0_\beta)}} }
\,,
\label{RMcn0}
\eea
\bea
n^0_{\alpha\beta}&\equiv&
\sum_{ij}c_ic_j\Tr\left(\phi_i^\dagger\phi_j^\dagga\,f^\dagger_{\alpha}f^\dagga_{\beta}\right)
\nonumber\\
&\equiv&\sum_{ij}c_ic_jN^{ij}_{\alpha\beta}
=\sum_{ij}c_ic_j\,\frac{1}{2}
\left(N^{ij}_{\alpha\beta}+N^{ji}_{\alpha\beta}\right)
\nonumber\\
&\equiv&\Av{c}{N^S_{\alpha\beta}}
\,,
\label{cfunc}
\eea
\bea
\Av{\Psi_0}{\mathcal{P}^\dagger\,\hat{H}_{\text{loc}}\,\mathcal{P}}&=&
\sum_{ij}c_ic_j\Tr\left(\phi_i^\dagger\,U\,\phi_j^\dagga\right)
\nonumber\\
&\equiv&\sum_{ij}c_ic_j\,U^{ij}\equiv\Av{c}{U}\,.
\eea

Notice that the tensors $M^S$, $N^S$ and $U$ 
are fully determined by the symmetry of the wavefunction~[Eq.~\eqref{GWFch1}] 
and the number of orbitals.
For this reason it is generally convenient to precalculate them before starting
the numerical minimization of the variational energy.
This point will be further discussed in Sec.~\ref{technicalremarks}.

\subsection{Simplified variational ansatz}\label{sva}
The complexity of the numerical problem is considerably reduced if 
the mixing of different atomic configurations is neglected in the 
Gutzwiller projector.
This amounts to assume that the matrix $\phi$ defined in Eq.~\eqref{phich1}
has the form
\bea
\phi&=&\sum_h c^{\text{int}}_k\,\phi_h^{\text{int}}\label{sva1}\\
\phi_h^{\text{int}}&\equiv&P_h^{\text{int}}/\sqrt{\Tr\left([P_h^{\text{int}}]^2\right)}
\label{sva2}\,,
\eea
where $P_h^{\text{int}}$ are the orthogonal projectors onto the eigenspaces of 
the local atomic interaction $\hat{H}_{\text{int}}$.
Although the variational parameters neglected in 
Eqs.~(\ref{sva1}-\ref{sva2}) can play a crucial role in some case,~\cite{lanata}
this simplified ansatz
merits to be mentioned for at least two reasons.  
(i) It still allows to solve exactly the problem in the atomic limit. 
(ii) The number of independent variational parameters is generally extremely lower 
in this approximation, allowing to perform calculations not feasible otherwise. 
Furthermore, once a variational result 
\be
\phi_0\equiv\sum_hc^{\text{int}}_{0\,h}\,\phi_h^{\text{int}}
\ee
is obtained assuming Eqs.~(\ref{sva1}-\ref{sva2}),
it can be used as a good starting point $c$
for the self-consistent search of the energy minimum with the more general 
variational space discussed before, see Eq.~\eqref{phikbasis},
\be
c_k=\Tr(\phi^\dagger_k\phi_0)\,,
\ee
with the result to speed up the calculation.
Notice, in fact, that $[\hat{H}_{\text{int}},G]=0$, implying that 
\be
[\phi_0,R(g)]=0\quad\forall g\in G\,,
\ee
i.e., that $\phi_0\in\mathcal{V}_\phi$.

\section{Numerical optimization of the variational energy}\label{enopt}

In this section we discuss in detail the self-consistent 
numerical strategy to minimize 
the energy~[Eq.~\eqref{variationalenergych1}] fulfilling the Gutzwiller 
constraints~[Eqs.~(\ref{uno-app-bisch1}-\ref{due-app-bis1ch1})].

Notice that the formulation of the Gutzwiller problem
through Eqs.~(\ref{uno-app-bisch1}-\ref{variationalenergych1})
is formally analog to the constrained formulation of DFT derived by
Levy~\cite{324,325} and Lieb.~\cite{327} In fact, the variational energy 
can be expressed as a functional of the \emph{variational} density 
matrix $n^0$, see Eq.~\eqref{vardm},
\be
\mathcal{E}_{\text{var}}[n^0]=\min_{n^0}
\mathcal{E}_{\text{var}}\left[c,\Psi_0\right]\,,
\label{dfn0}
\ee
where $\min_{n^0}$ denotes the minimum over the set of variational parameters
$c$ and $\ket{\Psi_0}$, satisfying the Gutzwiller 
constraints~[Eqs.~(\ref{uno-app-bisch1}-\ref{due-app-bis1ch1})]
at fixed $n^0$.
In this work Gutzwiller the problem is 
solved by calculating the density functional $\mathcal{E}_{\text{var}}[n^0]$, 
and minimizing it respect to $n^0$.

For clarity reasons we have structured the rest of this section as an
exposition of the numerical procedure, omitting the mathematical proofs. 
The mathematical details can be found in the appendices.

\subsection{Preliminary calculation}\label{translationxidai}


Let us consider the Gutzwiller renormalized non-local tight binding 
operator 
\bea
\hat{T}^G&=&\sum_{\alpha\beta}\sum_{\bR\neq\bRp} 
\tilde{t}^{\alpha\beta}_{\bR\bRp}\,\dc_{\bR\alpha}\da_{\bRp\beta}
\,, \label{tg}
\eea
where
\be
\tilde{t}_{\bR\bRp}^{\gamma\delta}\equiv\sum_{\alpha\beta}
t_{\bR\bRp}^{\alpha\beta}\,
\R_{\alpha\gamma} \R_{\beta\delta}\,.
\label{atilde}
\ee
For later convenience, we consider a general one-body Hamiltonian
\be
\hat{H}^G=\hat{T}^G+\Delta \hat{H}=\sum_{\bk n}\epsilon^G_{\bk n}\, 
\eta^\dagger_{\bk n}\eta^\dagga_{\bk n}\,,
\label{tgpdh}
\ee
where $\Delta \hat{H}$ is a given local operator.

Let $\ket{\Psi_0}$ be the ground state of $\hat{H}^G$.
It can be easily verified that 
\be
\frac{\partial\Av{\Psi_0}{\hat{T}^G[\R]}}{\partial \R_{\alpha\beta}}
=2\sum_{\bk} [{t}_{\bk} \R\, U_{\bk} f_{\bk}U^\dagger_{\bk}]_{\alpha\beta}
\label{finalderR}
\ee
where 
\bea
(f_{\bk})_{nm}&=&\theta(-\epsilon^G_{\bk n})\,\delta_{nm}
\label{trmateqs1}\\
\eta^\dagger_{\bk n}&=&\sum_i(U_\bk)_{\alpha n}\,\dc_{\bk \alpha}
\label{trmateqs2}\,.
\eea

Eq~\eqref{finalderR} will be used in the following subsections, 
where two important inner parts of our numerical scheme are described in detail.

\subsection{Slater determinant optimization step}\label{slatdetopt}

For later convenience, in this section we solve 
the problem to calculate the state $\ket{\Psi_0}$ that realizes the 
minimum of the Gutzwiller variational energy at fixed $c$ 
\bea
\E_{n^0,\R}\!\!&=&\!\!\min_{\ket{\Psi}\in S_{n^0}}\Av{\Psi}{\hat{T}^G}
\label{slatfunc}
\nonumber\\
S_{n^0}\!\!&\equiv&\!\!\left\{ 
\ket{\Psi}\,\text{t.c.}\;\Av{\Psi}
{\dc_{\bR\alpha}\da_{\bR\beta}}=\delta_{\alpha\beta}n^0_\alpha\right\},
\eea
where $\hat{T}^G$ is given by Eqs.~(\ref{tg}-\ref{atilde}).
Note that the functional $\E_{n^0,\R}$ depends on $c$ only indirectly
through $\R$, that is given by 
\be
\R_{\alpha\beta} = \frac{\Av{c}{{M}^S_{\alpha\beta}}}
{\sqrt{n^0_{\beta}(1-n^0_{\beta})}}\,,
\label{ttilderepeat}
\ee
see Eqs.~\eqref{Z-newch1} and \eqref{RMcn0}.

It is convenient to account for the Gutzwiller constraints employing the
Lagrange multipliers method. 
We introduce
\be
\hat{H}^G[\R,\lambda]=\hat{T}^G+\Delta\hat{H}\,,
\label{HGqp}
\ee
where
\be
\Delta\hat{H}=\sum_{\bR}\sum_{\alpha\beta}
\lambda_{\alpha\beta}\,\dc_{\bR\alpha}\da_{\bR\beta}\,.
\ee
Notice that $\hat{H}^G$ has the same form of Eq.~\eqref{tgpdh}.
Finally, we calculate the ground state $\ket{\Psi_0}$ of $\hat{H}^G[\R,\lambda]$
for $\lambda_{\alpha\beta}$ such that the Gutzwiller 
constraints~[Eqs.~(\ref{uno-app-bisch1}-\ref{due-app-bis1ch1})] 
are satisfied.
Once the Lagrange multipliers $\lambda_{\alpha\beta}$ are known we 
compute Eq.~\eqref{finalderR}.

In summary, the calculations described in this section associate the
input variables $n_\beta^0$ and $\R_{\alpha\beta}$ to the output matrix
\be
\mathcal{D}_{\alpha\beta}\equiv
\frac{\partial\Av{\Psi_0}{\hat{T}^G}}{\partial \R_{\alpha\beta}}\,,\label{defD}
\ee 
see Fig.~\ref{diagram}.

\subsection{$\phi$-matrix optimization step}\label{phimatoptsec}

In this section we derive the numerical procedure to minimize the 
Gutzwiller energy functional
\be
\mathcal{E}_{\Psi_0}[c]=\Av{\Psi_0}{\hat{T}^G}+\Av{c}{{U}}\,,
\label{variationalenergych1repeat}
\ee
where $\hat{T}^G$ is given by Eqs.~(\ref{tg}-\ref{atilde}) and
$\R$ depends on $c$ through Eq.~\eqref{ttilderepeat}, 
keeping the Slater determinant $\ket{\Psi_0}$ fixed and respecting the 
Gutzwiller constraints
\bea
\langle c|c\rangle &=& 1\label{uno-app-bisch1repeat}\\
\Av{c}{{N}^S_{\alpha\beta}} &=& 
\delta_{\alpha\beta}\,n^0_{\alpha}\,.
\label{due-app-bis1ch1repeat}
\eea

In order to solve this problem we adopt the following linearization 
procedure, that is based on appendix~\ref{philin}.
We consider the Hermitian matrix
\be
F[\mathcal{D},\lambda]=
H[\mathcal{D}]+{L}[\lambda]\,,
\label{eqstep2bis}
\ee
where $\mathcal{D}_{\alpha\beta}$ is defined by Eq.~\eqref{defD}, and
\bea
&&H[\mathcal{D}]
=U+\sum_{\alpha\beta}\mathcal{D}_{\alpha\beta}
\frac{{M}^{S}_{\alpha\beta}}{\sqrt{n^0_\beta(1-n^0_\beta)}}
\label{hd}\\
&&L[\lambda]=
\sum_{\alpha\beta}
\lambda_{\alpha\beta}\,{N}^{S}_{\alpha\beta}\,.\label{ll}
\eea
Then, we calculate the ground state $c$ of $F[\mathcal{D},\lambda]$ 
for $\lambda$ such that the Gutzwiller 
constraints~[Eqs.~(\ref{due-app-bis1ch1repeat}-\ref{due-app-bis1ch1repeat})] 
are satisfied. 
The obtained vector $c$ is used to define a new $\R$ through 
Eq.~\eqref{ttilderepeat}.

Notice that the matrix $\mathcal{D}_{\alpha\beta}$ entirely encodes
the dependency of the problem on $\ket{\Psi_0}$ in Eq.~\eqref{eqstep2bis}.
In summary, the above calculations associate to the
input variables $n_\beta^0$ and $\mathcal{D}_{\alpha\beta}$ the output
renormalizaton matrix $\R_{\alpha\beta}$, see Fig.~\ref{diagram}.

\subsection{Fixed point formulation}

\begin{figure}
  \centering
  \includegraphics{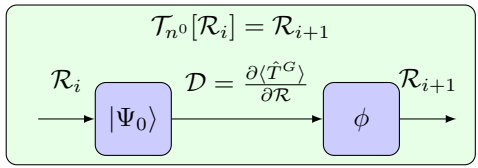}
  \caption{\label{diagram}(Color online) Flow chart representing the numerical calculation
    of the functional $\mathcal{T}_{n^0}$.}
\end{figure}

A very important observation in our implementation is that 
the composition of the two optimization steps derived 
in Secs.~\ref{slatdetopt} and \ref{phimatoptsec}
can be described as a functional $\mathcal{T}_{n^0}$ that associates a given
renormalization matrix $\R_{i}$ to a new renormalization matrix $\R_{i+1}$
\be
\R_{i+1}=\mathcal{T}_{n^0}[\R_{i}]\,,\label{forwarditeration}
\ee
see Fig.~\ref{diagram}. 
This operation lead to a reduction of the variational energy unless, 
by definition, $\R$ solves the equation
\be
\mathcal{T}_{n^0}[\R]-\R=0\,.\label{root}
\ee
In this case $\R$ defines a stationary point of the energy functional.
This observation amounts to formulate the minimization of the variational energy
at fixed $n^0$ as a \emph{fixpoint problem},
that can be solved in several ways.~\cite{mtheath}
A first possibility is to use $\R_{i}$ as an input to obtain $\R_{i+1}$ 
and iterate the procedure up to convergence.
This procedure is commonly referred to as \emph{forward recursion} 
method.~\cite{DMFT,borghi}
However, as it will be shown in Sec.~\ref{technicalremarks}, the application of the
\emph{Newton method} is generally much more efficient.

We underline that the size of $\R$ is equal to the number of orbitals,
and that the number of independent parameters that define it is often
reduced by symmetry.
For this reason the solution of Eq.~\eqref{root}
generally requires a few Newton steps to converge, see Sec.~\ref{technicalremarks}.
The problem of the exponential scaling of the local many-body space
affects the numerical algorithm exclusively through the solution for 
the ground state of $F[\mathcal{D},\lambda]$, see Eq.~\eqref{eqstep2bis}.
The size of the matrix $F[\mathcal{D},\lambda]$ is, in fact, equal to the dimension
of the vector $c$. 
Nevertheless, the 
calculation of the ground state of $F[\mathcal{D},\lambda]$ is not
numerically problematic for two reasons. 
(i) The dimension of $c$ is reduced by symmetries, 
as will be shown in Sec.~\ref{technicalremarks}.
(ii) The calculation of the ground state of $F[\mathcal{D},\lambda]$
does not require a full diagonalization.
Less computationally demanding algorithms, such as the power method or 
the Lanczos method, can be employed. See Sec.~\ref{technicalremarks}
for further details.

We remark that the self-consistent numerical algorithm derived in this paper 
requires, as a starting point, only a initial ``guess'' for 
the variational density matrix $n^0$ and the matrix $\R$. 
It is not necessary to construct a good 
initial guess for the whole Gutzwiller wavefunction, i.e., for the 
matrix $\phi$, while this would be necessary in order to perform 
a direct constrained minimization of the energy 
functional~[Eq.~\eqref{variationalenergych1}]. 
This implies that the stability of the algorithm is not affected by
the exponential scaling of the number of parameters involved in the 
calculation.
This point will be further discussed in Sec.~\ref{technicalremarks}.

\section{Application to LDA+Gutzwiller}\label{ldapg}

In this section we briefly discuss how to combine the Gutzwiller scheme with
a first principle calculation of the uncorrelated electron structure as 
an input, applying the DFT scheme with LDA. 
This combined scheme is named LDA+G.
We also outline how the Gutzwiller solver has to be modified in order to account for
the double counting.~\cite{LDA+U}
The double counting appears as a mean-field contribution in the exchange-correlation
taken into account in the LDA calculation.

As a starting point we consider the Kohn-Sham reference system obtained within
a converged LDA calculation
\bea
\hat{H}_{\text{LDA}}\!&=&\!\sum_{\bk}\sum_n\epsilon^{\textrm{KS}}_{\bk n}\,
\eta^\dagger_{\bk n}\eta^\dagga_{\bk n}\label{hks}\\
\eta^\dagger_{\bk n}|0\rangle\!&\equiv&\!\ket{\psi^{\textrm{KS}}_{\bk n}}\,.
\eea
In order to be able to apply LDA+G  
the tight binding Hamiltonian~[Eq.~\eqref{hks}] must first be expressed
in terms of a proper localized basis set~\cite{Fang}
$\ket{\chi_{k\alpha}}$, where $\alpha$ labels both spin $\sigma$ and 
orbital $a$. This can be done by means of the overlap matrix
\be
[S_{\bk}]_{n\alpha} = \langle\psi^{\textrm{KS}}_{\bk n}|\chi_{\bk \alpha}\rangle\,,
\ee
giving
\bea
\hat{H}_{\text{LDA}}\!&=&\!\sum_{\bk}\sum_{\alpha\beta}
\epsilon^{\alpha\beta}_{\bk}\,\cc_{\bk\alpha}\ca_{\bk\beta}\\
\epsilon^{\alpha \beta}_{\bk}\!&=&\!
\sum_n\, [S^\dagger_\bk]_{\alpha n}\,\epsilon^{\textrm{KS}}_{\bk n}\,[S_\bk]_{n\beta}\\
\cc_{\bk\alpha}|0\rangle\!&\equiv&\!\ket{\chi_{k\alpha}}
\,.
\eea

In order to express $\hat{H}_{\text{LDA}}$ in the same form of Eq.~\eqref{hamgen}
we separate $\hat{H}_{\text{LDA}}$ in a non-local part $\hat{T}$ and in a local part 
(the crystal fields) as follows
\bea
\hat{H}_{\text{LDA}}&=&\hat{T}+\hat{L}\\
\hat{T}&=&\sum_{\bk}\sum_{\alpha\beta}
t^{\alpha\beta}_{\bk}\,\cc_{\bk\alpha}\ca_{\bk\beta}\label{tlda}\\
\hat{L}&=&\sum_{\bR}\sum_{\alpha\beta}
l^{\alpha\beta}\,\cc_{\bR\alpha}\ca_{\bR\beta}\,,
\eea
where
\bea
l^{\alpha\beta}&=&\frac{1}{\Omega}\sum_{\bk}\epsilon^{\alpha\beta}_{\bk} \\
 t^{\alpha\beta}_{\bk}&=&\epsilon^{\alpha\beta}_{\bk}-l^{\alpha\beta}.
\eea
and $\Omega$ is the number of sites $\bR$. 
The on-site electron interaction can be modeled by the 
Slater-Kanamori rotationally invariant atomic interaction~\cite{Kanamori} 
\bea
\hat{H}_{\text{int}}&=&\sum_\bR\hat{H}^\bR_{\text{int}}\,,\\
\hat{H}^\bR_{\text{int}}&=&U\sum_a\hat{n}_{\bR a\uparrow}\hat{n}_{\bR a\downarrow}
+\frac{U'}{2}\sum_{a\neq b}\sum_{\sigma\sigma'}\hat{n}_{\bR a\sigma}\hat{n}_{\bR b\sigma'}
\nonumber\\
&-&\frac{J}{2}\sum_{a\neq b}\sum_{\sigma}
\cc_{\bR a\sigma}\ca_{\bR a-\sigma}\cc_{\bR b-\sigma}\ca_{\bR b\sigma}
\nonumber\\
&-&\frac{J'}{2}\sum_{a\neq b}
\cc_{\bR a\uparrow}\cc_{\bR a\downarrow}\ca_{\bR b\uparrow}\ca_{\bR b\downarrow}
\label{onsite}\,.
\eea
We underline that the form [Eq.~\eqref{onsite}] for $\hat{H}_{\text{int}}$ 
is obtained by implicitly assuming that 
the single-particle basis $\ket{\chi_{k\alpha}}$ is given by \emph{real} 
orbitals, i.e., the cubic (crystal) harmonics.
It can be proven that the condition $U=U'+J+J'$ ensures 
the rotational invariance in the orbital space.
The additional condition $J=J'$ can be assumed whenever 
the spin-orbital coupling is negligible.~\cite{Fang} 

Our model is now defined in the form of Eq.~\eqref{hamgen}, with
$\hat{T}$ given by Eq.~\eqref{tlda} and
$\hat{H}_{\text{loc}}=\hat{H}_{\text{int}}+\hat{L}$.
An additional on-site term, the double counting, needs to be added to 
Eq.~\eqref{variationalenergych1}
as the average orbital-independent interaction energy is already included in LDA. 
A common choice of the double counting term is~\cite{LDA+U,LDA+U+DMFT}
\bea
E_{\text{dc}}[\rho]
&=&\frac{\bar{U}}{2}\,n(n-1)
-\sum_{\sigma}\frac{\bar{J}}{2}\,n_{\sigma}(n_{\sigma}-1)\\
\bar{U}&=&\frac{U+2lJ}{2l+1}\nonumber\\
\bar{J}&=&\bar{U}-U'+J
\,,
\eea
where $l$ is the angular momentum quantum number of the considered localized 
basis set,
\bea 
n&\equiv&\sum_\sigma n_{\sigma}\equiv\sum_{a\sigma}n_{a\sigma}\nonumber\\ 
n_{\alpha}&\equiv&\rho_{\alpha\alpha}
\,,
\eea 
see Eq.~\eqref{ordm}, and $n_{\alpha}$
is the mean value of $\cc_{\bR\alpha}\ca_{\bR\alpha}$ with
respect to the Gutzwiller wavefunction, that
is given by
\bea
n_{\alpha}&=&\sum_{ij}c_ic_j\Tr\left(\phi_i^\dagger\,
\fc_{\alpha}\fa_{\alpha}\,\phi_j^\dagga\right)
\nonumber\\
&\equiv&\sum_{ij}c_ic_j\,P_{\alpha\alpha}^{ij}\equiv\Av{c}{P_{\alpha\alpha}}\,.
\eea
The presence of the double counting term gives rise to the following 
additional term in Eq.~\eqref{eqstep2bis}
\be
D[c]=\sum_\alpha
\frac{\partial E_{\text{dc}}}{\partial n_\alpha}P_{\alpha\alpha}\,.
\ee

When the self-consistent Gutzwiller calculation is converged the
result can be fed back to the LDA code.
By calculating the density matrix
\be
[\rho_{\bk}]_{\alpha\beta}\equiv
\Av{\Psi_0}{\mathcal{P}^\dagger\cc_{k\alpha}\!\ca_{k\beta}\mathcal{P}^\dagga\!\!}\,,
\ee
and representing it in the Kohn-Sham basis 
\be
[\rho_{\bk}^{\textrm{KS}}]_{nm}=\sum_{\alpha\beta}\,
[S_\bk]_{n\alpha}\,[\rho_{\bk}]_{\alpha\beta}\,[S^\dagger_\bk]_{m\beta}\,,
\ee
it is possible to get a prescription how to transform the Kohn-Sham eigenfunctions
and occupancies in order to reproduce the physical Gutzwiller electron 
density.~\cite{Fang}
To the new total density corresponds a new effective potential for the
Kohn-Sham reference system that, in turn,
defines a new tight binding Hamiltonian.

The procedure is iterated until self-consistency is reached.

\section{Illustrative results}\label{modres}

This section has two purposes. 
(i) As a proof of concept we present some numerical result in comparison
with other calculations based on different methods.
(ii) We outline several technical details of the calculations
and the speed of convergence of the algorithm.

Test calculations have been performed on multi-orbital model Hamiltonians
of the form
\bea
\hat{\h} &=& 
\sum_{\bk\sigma} \sum_{ab} t_{\bk}^{ab}
\cc_{\bk a\sigma}\ca_{\bk b\sigma} \nonumber\\
&+& \sum_{\bR\sigma}\sum_{ab}
l^{ab}\,\cc_{\bR a\sigma}\ca_{\bR b\sigma} + \hat{H}_{\text{int}}\,,
\label{modelham}
\eea
where the structure of $\hat{H}_{\text{int}}$ was defined in Eq.~\eqref{onsite}. 
A paramagnetic Gutzwiller wavefunction has been assumed in all the calculations
shown in this section. 
The hopping matrix has
been set up as either (i) nearest neighbor hopping on a three
dimensional cubic lattice,
giving a non-interacting DOS with cusps close to half the bandwidth,
or (ii) nearest neighbor hopping on a Bethe
graph with infinite coordination number, corresponding to a semicircular
non-interacting DOS. In both cases the half bandwidth $W$ is set as
the unit of energy.

\subsection{Two-bands Hubbard model}\label{res2}

\begin{figure}
  \centering
  \includegraphics{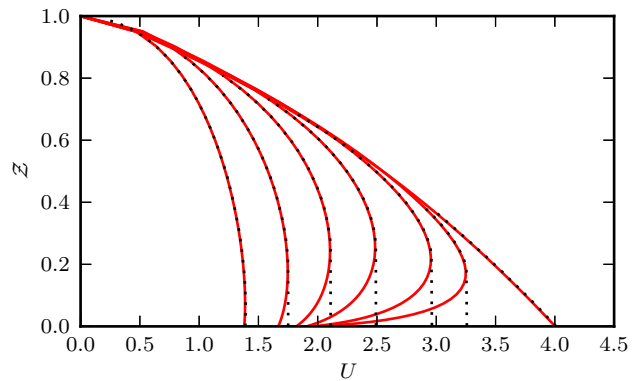}
  \caption{\label{3DCubicZ}(Color online) Comparison of results, for two
    degenerate bands on the 3D cubic lattice with rotational-invariant Hunds interaction, 
    from Gutzwiller (solid lines) and slave-boson~\cite{Lechermann:2007ys} (dotted
    lines), showing the quasi-particle weight $\mathcal{Z}$ as
    a function of $U$ and (from right to left), $J/U = 0$, $0.01$, $0.02$, $0.05$, $0.10$,
    $0.20$, $0.45$.
    Additionally, also the higher energy fixpoint solutions of 
    Eq.~\eqref{root} at finite $J/U$ are reported.}
\end{figure}

First of all, let us consider the case of two orbitals.
In the special case of half-filling and degenerate bands 
we compare our calculations with the available results obtained 
from Ref.~\onlinecite{Lechermann:2007ys}
by means of the rotationally invariant slave-boson 
technique,~\cite{Barnes1,Barnes2,Coleman-1/N,Read&Newns,Lechermann:2007ys}
that is equivalent to the Gutzwiller variational method 
on the mean field level.~\cite{Kotliar-Ruckenstein,quaquaraqua}
In this case $t_{\bk}^{ab}$ is set up as nearest neighbor hopping 
on a three dimensional cubic lattice
\be
\label{degbands}
t_\bk^{ab}=-\,t^0_{ab}\,\frac{1}{3}\sum_{\mu=1}^3\cos(k_\mu)\,,
\ee
with $t^0_{ab}=\delta_{ab}$.
For this specific model, in which the single-particle energy dispersion 
of the two bands are identical, we have that
\be
\R_{ab}=\sqrt{\mathcal{Z}}\,\delta_{ab}\,,\label{rz}
\ee
where $\mathcal{Z}$ can be interpreted as a measure of the
quasi-particle renormalization weight.~\cite{Gebhard-FL} 
As shown in Fig.~\ref{3DCubicZ}, the Gutzwiller calculation gives
the same values of $\mathcal{Z}$ as a function of $U$ and $J/U$ as the 
slave-boson calculations.
The Brinkman-rice transition~\cite{brinkman&rice} occurs at 
a strongly $J$-dependent critical $U$. 
This well known fact~\cite{PhysRevB.66.165107}
supports the argument that the spin-exchange on-site interaction 
needs to be taken into account to accurately describe strongly correlated systems.
Furthermore, while for $J/U=0$ the phase transition is second order, at finite $J/U$
it becomes first order, with a hysteresis region 
characterized by an additional fixpoint solution of Eq.~\eqref{root},
corresponding to a second stationary point of the variational energy,
see Fig.~\ref{3DCubicZ}.
To our knowledge this solution has never been reported before neither in the Gutzwiller 
nor in the slave-boson approximation.

Let us consider the more general case of non-degenerate bands
with finite crystal field splitting $\Delta$ 
\be
\sum_{ab}l^{ab}\,\cc_{\bR a\sigma}\ca_{\bR b\sigma}=
\Delta\,\left(\hat{n}_{\bR 1\sigma}-\hat{n}_{\bR 2\sigma}\right)\,.
\ee
This model has been studied in detail 
with DMFT in Ref.~\onlinecite{Werner:2007lr}.  
Here we compare our Gutzwiller results with part of the available DMFT data.
This gives an opportunity to discuss some features of the 
specific implementation derived in this work and to introduce some
general merits and limits of the Gutzwiller variational method in itself.

\begin{figure}
  \centering
  \includegraphics{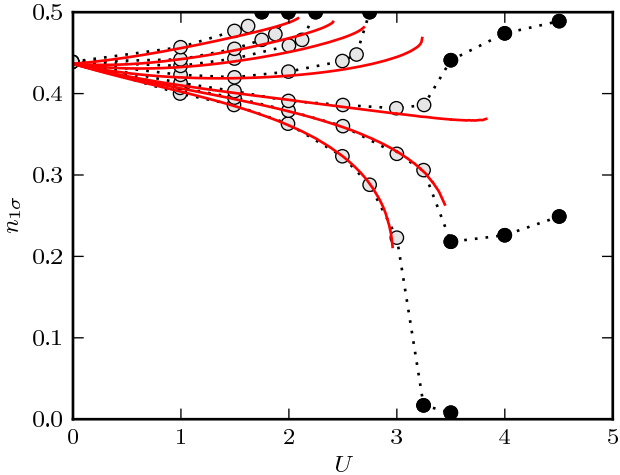}
  \caption{\label{n1GUTZandDMFT}(Color online)  
    Filling per spin of orbital $1$ 
    for $\Delta=0.2$ and different values of $J/U$. From bottom to top, 
    $J/U=0, 0.01, 0.02, 0.05, 0.1, 0.15, 0.25$.
    The continue lines correspond to our Gutzwiller results for the metallic 
    phase at half-filling. 
    Open (full) symbols correspond to metallic (insulating) solutions 
    obtained from DMFT~\cite{Werner:2007lr} at inverse temperature $\beta=25$.
  }
\end{figure}

In Fig.~\ref{n1GUTZandDMFT} the expectation value of the filling 
per spin $\hat{n}_{1\sigma}$ is shown for several values of $U$ and $J/U$.
Following Sec.~\ref{phiform}, the expectation value of $\hat{n}_{1\sigma}$
is given by
\be
n_{1\sigma}=\Tr(\phi^\dagger\,f^\dagger_{1\sigma}f^\dagga_{1\sigma}\,\phi)\,,
\ee
see Eq.~\eqref{avloc}.
In the same figure the DMFT results from Ref.~\onlinecite{Werner:2007lr}
are also shown.
These calculations were performed assuming a semicircular density of states. 

For the specific crystal field splitting considered, $\Delta=0.2$,
the system is metallic at small $U$,
and is driven towards a Mott insulating or an orbitally polarized phase,
depending on the value of $J/U$,
upon increasing the interaction strength $U$. 
Notice that the DMFT and the Gutzwiller results are in very good agreement
in the metallic phase, especially away from the Mott transition. 
This confirms that the quality of the Gutzwiller calculations is generally 
comparable with the quality of DMFT for the ground state properties of strongly 
correlated metals.~\cite{Fang}
On the contrary, the Mott insulating phase can not be described correctly
by means of a Gutzwiller approximation.
Nevertheless, it is correct to assume that $\mathcal{Z}$ approaching 
$0$ indicates that the metallic phase becomes unstable,
compatibly with the Brinkman-rice scenario.~\cite{brinkman&rice}
Notice that the critical coupling $U_c(J)$ of the Mott transition
predicted by the Gutzwiller approximation is not accurate in general.
For instance, at $J=0$ the Gutzwiller calculation gives
$U_c\approx 5$ for the semicircular density of states,
which differs from the DMFT result~\cite{Werner:2007lr} 
$U_c^{\textrm{DMFT}} \approx 4$ by $20\%$ (not shown).
Notice that the results of Fig.~\ref{3DCubicZ} were obtained 
assuming a three dimensional cubic lattice, and not a semicircular 
density of states.

\subsection{Five-bands Hubbard model}\label{res5}

In this section we study the Hamiltonian of the form~[Eq.~\eqref{modelham}]
for five bands, describing correlated $d$-electrons in a cubic crystal.
We assume a semicircular density of states, and that
the full rotational symmetry is broken by a finite crystal field splitting
$\Delta$ between the three $t_{2g}$ and the two $e_g$ orbitals, i.e.,
\be
\sum_{ab}l^{ab}\,\cc_{\bR a\sigma}\ca_{\bR b\sigma}=
\Delta\,\sum_{a\in t_{2g}}\hat{n}_{\bR a\sigma}\,.
\ee
This model has previously been used as a benchmark system in DMFT.~\cite{Lauchli:2009fk}
In our calculation we assume a paramagnetic Gutzwiller wavefunction 
invariant with respect to the symmetry point group of the cube.
This allows to reduce considerably the number of variational parameters,
see Sec.~\ref{technicalremarks}.

\begin{figure}
  \centering
  \includegraphics{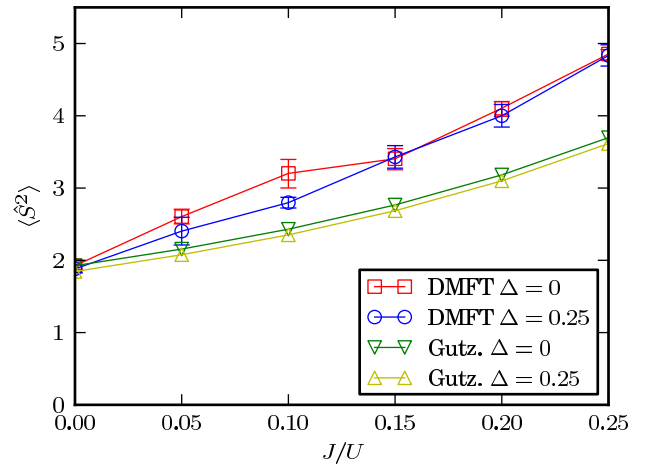}
  \caption{\label{S2GUTZandDMFTfive}(Color online)   
    Gutzwiller expectation value of $\hat{S}^2$
    for different values of $J/U$ and crystal field splittings
    $\Delta$, the total filling is $6$ electrons per site and $U = 1$. 
    Comparison with DMFT data 
    at inverse temperature $\beta=25$ from Ref.~\onlinecite{Lauchli:2009fk}.
  }
\end{figure}

Let us consider the expectation value $S^2$ of the total spin squared $\hat{S}^2$,
for which DMFT data are available in Ref.~\onlinecite{Lauchli:2009fk}.
From Eq.~\eqref{avloc} we have that 
\be
S^2=\Tr(\phi^\dagger\,\mathbf{S}^2\,\phi)\,,
\ee
where
\be
\mathbf{S}_k=\sum_{a}\sum_{\sigma\sigma'}
f^\dagger_{a\sigma}\frac{\sigma^k_{\sigma\sigma'}}{2}f^\dagga_{a\sigma'}
\ee
and $\sigma^k$ are the Pauli matrices.
In Fig.~\ref{S2GUTZandDMFTfive} the behavior of $S^2$  is shown at fixed $U=1$ 
for $6$ electrons per site and several values of $J/U$.
In the figure our results are compared 
with the DMFT data from Ref.~\onlinecite{Lauchli:2009fk}.
Consistently with DMFT, we find that $S^2$ grows monotonically upon 
increasing $J/U$, and that  
the crystal field splitting $\Delta=0.25$ slightly reduces $S^2$
compared to the case of degenerate bands. 
Notice that the discrepancy between the Gutzwiller results and the 
DMFT data becomes larger upon increasing $J/U$ at fixed $U$. 
A similar qualitative behavior could be observed even in the calculation
shown in Fig.~\ref{3DCubicZ}
for the two-bands model.
Nevertheless, the observed deviation between the Gutzwiller 
results and the DMFT data seems to be more substantial in this case.

\subsection{Bilayer Hubbard model}\label{resbil}

In both the models previously considered the renormalization
matrix $\R$ was diagonal due to symmetry. 
For completeness, we also consider the 
bilayer Hubbard model,~\cite{fab,Lechermann:2007ys,mybil}
in which $\R$ have finite off-diagonal elements.
In particular, we consider the Hamiltonian given by Eq.~\eqref{modelham}
assuming that the local hybridization term described by the matrix
\be
l=\left(
\begin{array}{cc}
0 & V \\
V  & 0 
\end{array}
\right)
\label{general_l}
\ee
with $V=0.25$, 
and a hopping matrix $t_\bk^{ab}$ given by Eq.~\eqref{degbands}
as in Sec.~\ref{res2}.
Finally, we assume that the local interaction $\hat{H}_{\text{int}}$
is given by Eq.~\eqref{onsite} with $U'=J'=J=0$.
This model has previously been studied -- with the same parameters --
in Ref.~\onlinecite{Lechermann:2007ys} with the slave-boson method. 

When the bandwidths are equal for the two bands (as in the present case) 
the matrix $l$ defined in Eq.~\eqref{general_l} can be diagonalized without 
modifying the hopping matrix $t_\bk^{ab}$ for any $\bk$.~\cite{Lechermann:2007ys}
This change of basis transforms the hybridization term $l$ in a crystal field
splitting between the bonding (+) and antibonding (-) orbitals. 
In the new basis both the effective renormalization matrix 
$\mathcal{Z}^0\equiv[\R^0]^2$ and the density matrix are diagonal
\be
\mathcal{Z}^0=\left(
\begin{array}{cc}
\mathcal{Z}_+ & 0 \\
0   & \mathcal{Z}_- 
\end{array}\right)\,,\;
n^0=\left(
\begin{array}{cc}
n_+ & 0 \\
0   & n_- 
\end{array}\right)\,,
\ee
and the coefficients $\mathcal{Z}_+,\mathcal{Z}_-$ 
can be interpreted as the quasi-particle renormalization 
weights of the bonding and antibonding orbitals respectively.
In the original basis, instead, $\R$ has non-zero off-diagonal elements and
$\mathcal{Z}\equiv\R^2$ has the 
form $\mathcal{Z}_{11}=\mathcal{Z}_{22}$ and 
$\mathcal{Z}_{12}=\mathcal{Z}_{21}$, where
\bea
\mathcal{Z}_{11}&=&\frac{\mathcal{Z}_++\mathcal{Z}_-}{2}\label{zd}\\
|\mathcal{Z}_{12}|&=&\frac{|\mathcal{Z}_+-\mathcal{Z}_-|}{2}\label{zoffd}\,.
\eea

\begin{figure}
  \centering
  \includegraphics{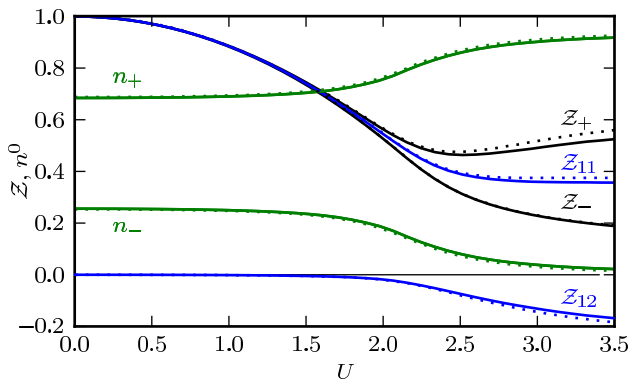}
  \caption{\label{2band_LocalHyb_SB_cmp}(Color online)   
    Gutzwiller renormalization matrix $\mathcal{Z}$ and filling of the bonding-antibonding
    bands for the two-bands bilayer Hubbard model with equal bandwidths,
    local hybridization $V=0.25$, $U'=J'=J=0$,
    and filling $N = 1.88$ (solid lines). 
    Comparison with slave-boson
    results from Ref.~\onlinecite{Lechermann:2007ys} (dotted lines). 
  }
\end{figure}

In order to compare with the slave-boson results of Ref.~\onlinecite{Lechermann:2007ys}
we have studied the system for $N=1.88$ electrons per site.
As seen in Fig.~\ref{2band_LocalHyb_SB_cmp},  the average of 
$\mathcal{Z}_+$ and $\mathcal{Z}_-$, given by $\mathcal{Z}_{11}$,
decreases monotonically as a function of $U$ as expected.
Concomitantly, the difference between $\mathcal{Z}_+$ and $\mathcal{Z}_-$,
given by $|\mathcal{Z}_{12}|$, progressively increases with $U$.
Our calculations compare well with the slave-boson results, although we find a 
slightly lower renormalization of the antibonding state at large interactions.

\subsection{Technical remarks}\label{technicalremarks}

\begin{figure}
  \centering
  \includegraphics{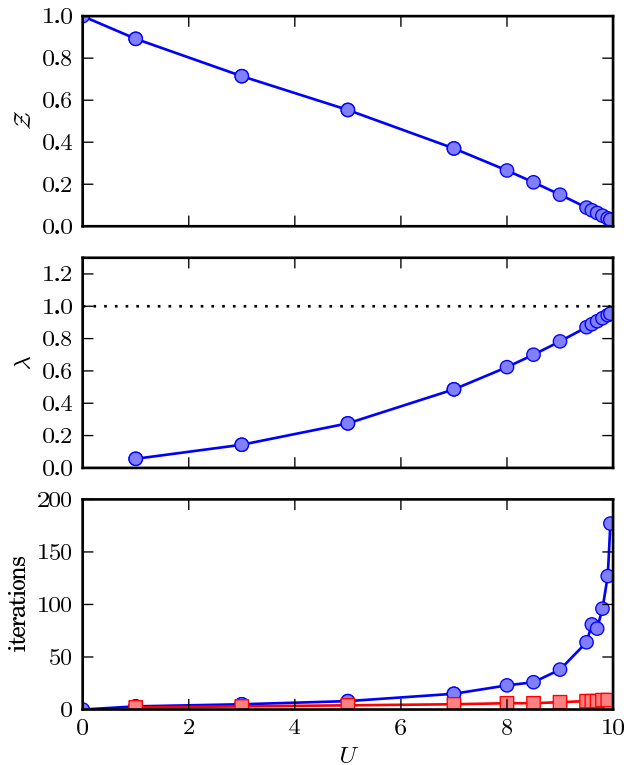}
  \caption{\label{5bands_convergence_study}(Color online)   
  Convergence of the forward recursion scheme (circles) and of the Newton method 
  (squares) for 5 bands at $N=5$, $J=0$ and $\Delta = 0$. 
  The quasi-particle weight $\mathcal{Z}\equiv\R_{\alpha\alpha}^2$ goes to zero 
  as $U$ goes to the critical coupling $U_c \approx 10\,$. 
  Simultaneously, the leading eigenvalue $\lambda$ of the
  Jacobian of the recursion function goes to $1$, and the number of
  forward iterations required to reach a fixed relative precision 
  (here $|\R_i-\R_{i+1}| \le 10^{-6}$ is used) diverges.
  On the contrary, the number of Newton iterations is almost independent
  of $U$.
  The matrix $[\R_0]_{\alpha\beta}=\delta_{\alpha\beta}$ is used 
  as initial condition of both the forward recursion series and the 
  Newton method $\forall U$.
  }
\end{figure}

In this section we point out several technical details of the 
numerical simulations performed to derive the results presented above.

The main technical problem of the Gutzwiller method is that 
the dimension of $c$, see Eq.~\eqref{phikbasis},
scales exponentially with the number of correlated orbitals. 
Fortunately, this number can be highly reduced by taking into account the symmetries 
of the system.
As an example, the number of matrix elements of $\phi$
(i.e., the dimension squared of the local Fock space)
is compared in table~\ref{tab:Dimensions} with the dimension of the vector $c$
in the case of a paramagnetic Gutzwiller wavefunction invariant with respect to 
the point symmetry group of the cube.
This simplification is very important, as $\text{dim}(c)$ is equal to the size of 
$F[\mathcal{D},\lambda]$, see Eq.~\eqref{eqstep2bis}, whose ground state
needs to be evaluated many times during the calculations.
To compute the ground state of $F[\mathcal{D},\lambda]$, i.e., 
its eigenvector with the lowest eigenvalue, 
we have used the iterative Arnoldi based solver provided by the
ARPACK library.   
This calculation is further speeded up by exploiting the sparsity 
of $F[\mathcal{D},\lambda]$,
effectively reducing the cost of the necessary matrix-vector multiplications.


As anticipated in Sec.~\ref{tensors}, it is generally convenient to precalculate
 $\phi_k$ and the tensors $M^{ij}_{\alpha\beta}$, $N^{ij}_{\alpha\beta}$ and $U^{ij}$ in order to further speed up the calculations. This reduces the construction of the matrix $F$ to the sums in Eqs.~(\ref{hd}-\ref{ll}), but increases the memory requirements. In fact, the total number of elements $N_T$ in the tensors scales as,
\be
N_T = (2 N_{\text{orb}}^2 + 1) N_c^2  + N_c N_\Gamma^2\,,
\ee
where $N_c$ is the dimension of the vector $c$, $N_{\text{orb}}$ is the number of 
orbitals, and $N_\Gamma=2^{2N_{\text{orb}}}$ is the dimension of the local space.
However, the number of stored matrix elements can considerably reduced by exploiting 
the sparsity of the tensors.  
For instance, the number of non-zero elements in the tensors was reduced by 
around three orders of magnitude for the five-bands Hubbard model of 
Sec.~\ref{res5}.
Eventually, in more complicated calculations it may happen that 
the number of variational parameters is so large that 
the tensors can not be stored in memory.
In this case it is still possible to calculate the 
traces ``on the fly''. This operation is trivially parallelizable.


It is well known that the speed of convergence of the forward recursion 
method~[Eq.~\eqref{forwarditeration}] is limited by the magnitude of the largest
eigenvalue $\lambda$ of the Jacobian of the transformation 
$\mathcal{T}_{n^0}$ in the fixed point $\R$.
In particular, if $|\lambda|\rightarrow 1$
the rate of convergence displays a critical slowing down.
We have found that this situation actually occurs in our simulation
when $U$ approaches the Brinkman-Rice critical value.
This is shown in Fig.~\ref{5bands_convergence_study}, 
where the convergence of $\R$ 
is shown for different values of $U$ at half-filling ($N=5$), $J=0$ 
and $\Delta=0$. The value of $\lambda$ was obtained as 
\bea
\lambda &=& \lim_{i\rightarrow\infty}(\R_{i+1}-\R)/(\R_{i}-\R)\nonumber\\
\R&=& \lim_{i\rightarrow\infty}\R_i\,,
\eea
where $\R_i$ was obtained from the forward recursion 
series~[Eq.~\eqref{forwarditeration}] starting from the initial condition
\be
[\R_0]_{\alpha\beta}=\delta_{\alpha\beta}\quad\forall U\,.
\ee
In DMFT the self energy $\Sigma$ is obtained as the solution
of a fixed point problem,~\cite{DMFT}
analogously to the matrix $\R$, that is the solution of Eq.~\eqref{root}.
It is known that also in DMFT the rate of convergence of the forward recursion method
displays a critical slowing down in the vicinity of the Mott transition.
This behavior has recently been shown to 
be related to the fact that the maximum eigenvalue of the Jacobian
$\lambda$ approaches $1$ as $U$ approaches the critical value~\cite{hugo},
as in our Gutzwiller calculations.
In this case the convergence problem has been cured by
employing quasi-Newton methods instead of the forward recursion scheme.~\cite{zitko,hugo}
The same strategy is applicable to solve Eq.~\eqref{root}.
As shown in Fig.~\ref{5bands_convergence_study}, this strategy is very efficient.
While the forward recursion algorithm slows down
as $U$ approaches its critical value,
the number of required Newton iterations is essentially independent of $U$.
The time required to calculate the results shown in 
Fig.~\ref{5bands_convergence_study} with the Newton method is less than one minute
for every single $U$.
Nevertheless, the forward recursion method could be more stable in some case, as 
every forward recursion step leads to a decrease in total energy. 
In fact, see Sec.~\ref{enopt}, every evaluation of $\mathcal{T}_{n^0}$
corresponds to a minimization of the energy 
with respect to the Slater determinant followed by a minimization with respect 
to the Gutzwiller projector. 
This guarantees that the fixed points calculated by the forward-recursion method
are local minima, while the Newton method can converge also to fix points 
with one or more Jacobian-eigenvalues $|\lambda|>1$, i.e., to stationary points 
of the energy that are not local minima.

\begin{figure}
  \centering
  \includegraphics{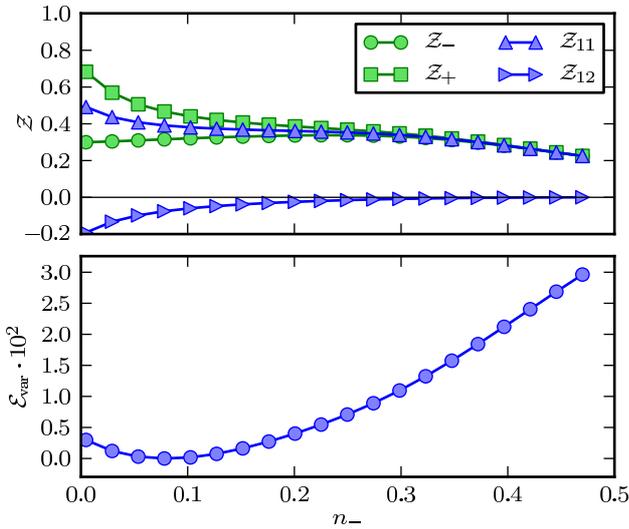}
  \caption{\label{2band_LocalHyb_U2_5_psweep}(Color online)   
    Sweep in the filling of the antibonding orbital $n_{-}$ for 
    the two-bands bilayer Hubbard 
    model with equal bandwidths, local hybridization $V=0.25$, 
    filling $N = 1.88$, $U'=J'=J=0$ and $U=2.5$.
    Renormalization matrix $\mathcal{Z}$ (top panel) and 
    total energy (bottom panel). 
  }
\end{figure}

We remark that the numerical procedure proposed in this paper is divided 
in two steps.
(i) Construction of the functional $\mathcal{E}_{\text{var}}[n^0]$ 
optimizing the variational energy for a fixed variational density matrix $n^0$. 
This optimization can be reduced to the fixed point problem~[Eq.~\eqref{root}]
and solved with the methods discussed above.
(ii) Direct minimization of $\mathcal{E}_{\text{var}}[n^0]$ with respect to $n^0$.
For completeness, this procedure is illustrated explicitly here
for the bilayer Hubbard model discussed in Sec.~\ref{resbil}.
For each value of $U$ the total energy $\mathcal{E}_{\text{var}}[n^0]$ was 
optimized with respect to the variational density $n^0$ using a bound 
minimization routine.  
An example is shown in Fig.~\ref{2band_LocalHyb_U2_5_psweep},
where the renormalization factors and the total energy are shown for fixed $U=2.5$
and total filling per site $N=1.88$
as a function of the antibonding orbital filling $n_{-}$. 
For each $n_{-}$ the fixpoint problem of Eq.~(\ref{root}) was solved employing a 
quasi-Newton method with convergence criterion 
$|\R_i-\R_{i+1}| \le 10^{-12}$ in less than 40 steps.

Finally, we point out
that the presence of finite off-diagonal terms in $\R$ is due to
the generality of the variational ansatz considered in this work. 
In fact, $\R$ is always diagonal if
Eqs.~(\ref{ppp1}-\ref{ppp2}) hold, as it was assumed in Ref.~\onlinecite{Fang}, 
see Sec.~\ref{dprojcase}.

\begin{table}
  \caption{\label{tab:Dimensions}
    Dimensions of the variational space for 1, 3 and 5 atomic orbitals
    (corresponding to $s$, $p$ and $d$ electrons). 
    The number of matrix elements of $\phi$, $2^{4N_{\text{orb}}}$,
    is compared with the dimension of the vector $c$, that
    is reduced by the point group symmetry of the cubic lattice.}
  \begin{ruledtabular}
    \begin{tabular}{lccrr}
      & $l$ &  $N_{\text{orb}}$ &$2^{4N_{\text{orb}}}$ & $\text{dim}(c)$ \\
      \hline
      s & 0 & 1 & 16  &  3 \\
      p & 1 & 3 & 4096 & 16  \\
      d & 2 & 5 & 1048576  &  873\\
    \end{tabular} 
  \end{ruledtabular}
\end{table}

\section{Conclusions}\label{concl}
In this article we have derived a numerically efficient self-consistent 
implementation of the Gutzwiller variational method. 
The method proposed was obtained as a combination of 
the self-consistent numerical procedure recently derived by Deng \emph{et al.} 
in Ref.~\onlinecite{Fang} and the mathematical formulation of the Gutzwiller problem 
developed by Fabrizio and collaborators.~\cite{Attaccalite,ferreroPRB,fab,lanata,mybil,mythesis}
This formalism allows us to overcome the restriction to density-density 
interaction, that was assumed in Ref.~\onlinecite{Fang}, without increasing the 
complexity of the numerical algorithm. 
The approach drastically reduces the problem of the high-dimensional Gutzwiller minimization
by mapping it to a minimization only in the variational density matrix, in the spirit of the
Levy~\cite{324,325} and Lieb~\cite{327} formulation of DFT. 
For fixed density the Gutzwiller renormalization matrix 
is determined as a fixpoint of a proper functional of $\R$,
whose evaluation only requires ground-state calculations
of matrices defined in the Gutzwiller variational space. We have compared
different methods to solve the fixpoint problem, finding that the Newton method 
is generally more efficient than the forward iteration method. 
The formalism also allows us to reduce the number of independent variational parameters
in a well controlled way using symmetries.
As a proof of concept we have performed a few numerical
calculations for two and five band Hubbard models with full rotationally invariant
interaction, finding good agreement with available DMFT and slave-boson data.
This analysis shows that the numerical approach derived is very stable and efficient.
For these reasons the scheme is promising 
for first-principle studies of real materials, e.g., in combination with DFT (LDA+G).

It is noteworthy that the numerical implementation presented in this work allows
for straightforward extensions in two directions of interest.
(i) The variational freedom of the Gutzwiller wavefunction can be generalized
in order to describe superconducting~\cite{fab,mybil} and magnetic 
systems.~\cite{lanata}
(ii) The assumption that the coefficients $\lambda$ of the 
Gutzwiller projector~[Eq.~\eqref{PRch1}] are real can be dropped, 
and generalized to complex values.
This allows, for instance, to account also for spin-orbit
corrections to the on-site interaction.

\begin{acknowledgments}

We are grateful to Michele Fabrizio and Giovanni Borghi for discussions.
We also thank Frank Lechermann for providing us with with the slave-boson data 
in Figs.~\ref{3DCubicZ} and \ref{2band_LocalHyb_SB_cmp},
and Philipp Werner for providing us with with the 
DMFT data in Figs.~\ref{n1GUTZandDMFT} and~\ref{S2GUTZandDMFTfive}.

We acknowledge funding from the Mathematics-Physics Platform (MP$^2$)
at the University of Gothenburg.
The simulations were performed on resources provided by the Swedish
National Infrastructure for Computing (SNIC) at Chalmers Centre for
Computational Science and Engineering (C3SE)
(project 001-10-37).

\end{acknowledgments}

\appendix

\section{Irreducible representation for a paramagnetic wavefunction}\label{appsym}

In this appendix we explain how to calculate the transformation $V$ 
introduced in Eq.~\eqref{transftobar} for a general group $G$. 
Let us consider the example in which $G$ is the symmetry group of a 
paramagnetic wavefunction invariant with respect to a specific discrete group
of real (orbital) rotations $G_{\text{orb}}$, 
e.g., the symmetry group of the cube. 
The problem consists in the definition of the most general $\phi$ matrix that 
commutes with the number operator $\hat{N}$, the representation of the 
spin $\hat{\mathbf{S}}$ and the representation $\hat{G}_{\text{orb}}$ 
of $G_{\text{orb}}$.
\bea
[\phi,\hat{N}]=
[\phi,\hat{\mathbf{S}}]
=[\phi,\hat{g}]&=&0\quad \forall\hat{g}\in\hat{G}_{\text{orb}}
\,.\label{symgroup}
\eea

In order to achieve our purpose the first step is to diagonalize 
simultaneously $\hat{N}$ and $\hat{\mathbf{S}}^2$.
Each simultaneous eigenspace of these operators
is the basis of a representation of the symmetry group $G$ identified
by the eigenvalues $(N,S)$. We denote such a space $V_{N,S}$.
Let us decompose $V_{N,S}$ in irreducible representations
of the spin rotations. 
In order to do this we consider the kernel of the spin lowering operator $\hat{S}_-$
in $V_{N,S}$, and denote it by $V_{N,S,-S}$.
Then, we calculate an orthonormal basis of $V_{N,S,-S}$ 
in which, for later convenience,
also $\hat{\mathbf{L}}^2$ and $\hat{L}_Z$ are diagonal
\be
V_{N,S,-S}=\text{Span}\left(\{\psi^{L,m_L,i}_{N,S,-S}\}\right)\,. 
\ee
To each value of $L$, $m_L$ and $i$ corresponds, by applying to $\psi^{L,m_L,i}_{N,S,-S}$ 
the raising $\hat{S}_+$ operator up to $2s$ times, 
a set of states labeled as $\{\psi^{L,m_L,i}_{N,S,m_S}\}$.
Each subset $V^{L,m_L,i}_{N,S}$ is defined as
\be
V^{L,m_L,i}_{N,S}=\text{Span}\left(\{\psi^{L,m_L,i}_{N,S,m_S}\,|\,m_s=-S,..,S\}\right)\,,
\ee
is a basis of an irreducible representation of the spin group.

It is clear that, in order to commute with $\hat{N}$ and $\hat{\mathbf{S}}^2$, 
$\phi$ is decomposed in uncoupled blocks, each of them acting on 
the corresponding subspace $V_{N,S}$. 
In each block we group together the vectors $\{\psi^{L,m_L,i}_{N,S,m_S}\}$ 
with the same $m_S$.
The Schur lemma~\cite{Wigner} ensures that, if the above order convention
is used, the $(N,S)$ block of $\phi$ has the general form
\be
\left.\phi\right|_{V_{N,S}}=
\left(
\begin{array}{ccc}
p^{N,S} & \cdots\ & 0 \\
\vdots  & \ddots & \vdots \\
0 & \cdots & p^{N,S}  \\
\end{array}
\right)\,.
\label{formphiblock}
\ee

The structure of the matrix $p$ in Eq.~\eqref{formphiblock} is 
further reduced by the condition $[\phi,\hat{G}_{\text{orb}}]=0$.
The vector space $V_{N,S,m_S}$ generated by 
$\{\psi^{L,m_L,i}_{N,S,m_S}\}_{L,m_L,i}$
is the basis of a representation of $G_{\text{orb}}$, 
that can be decomposed in irreducible representations using standard methods. 
Notice, in fact, that a state $\psi^{L,m_L,i}_{N,S,m_S}$ transforms exactly as 
the spherical harmonic function $Y^L_{m_L}$ under rotations.
Each one of the obtained irreducible representation of $G_{\text{orb}}$ 
is labeled by its characters.~\cite{Wigner} 
We group together all the equivalent representations
with equal characters $\chi$. 
This amounts to express $V_{N,S,m_S}$ as the direct sum of $V_{N,S,m_S}^\chi$.
The Schur lemma~\cite{Wigner} ensures that each one of the 
$p^{N,S}$ blocks defined in Eq.~\eqref{formphiblock} has the general form
\be
p^{N,S}=
\left(
\begin{array}{ccc}
q^{N,S}_{\chi_1} & \cdots\ & 0 \\
\vdots  & \ddots & \vdots \\
0 & \cdots & q^{N,S}_{\chi_{n_{\text{ch}}}}  \\
\end{array}
\right)\,,
\label{formphiblock2}
\ee
where $n_{\text{ch}}$ is the number of inequivalent representation in $V_{N,S,m_S}$.

The final step is to identify the $n_k$ states of each subspace
$V^{\chi_k}_{N,S,m_S}$ 
that belong to the same \emph{row}~\cite{Wigner} of the corresponding
(equivalent) irreducible representations of $G_{\text{orb}}$. 
The Schur lemma~\cite{Wigner} 
restricts the structure of each $q^{N,S}_{\chi_k}$ block of Eq.~\eqref{formphiblock2} 
as in Eq.~\eqref{formphi_V_equivalent}, i.e.,
\be
q^{N,S}_{\chi_k} = \left(
\begin{array}{ccc}
r^{N,S,\chi_k}_{11}\mathbbm{1}_{d_k} & \cdots & r^{N,S,\chi_k}_{1 n_k}\mathbbm{1}_{d_k} \\
\vdots  & \ddots & \vdots \\
r^{N,S,\chi_k}_{n_k 1}\mathbbm{1}_{d_k} & \cdots & r^{N,S,\chi_k}_{n_k n_k}\mathbbm{1}_{d_k} \\
\end{array}
\right)
\,,
\label{formphiblock3}
\ee
where $\mathbbm{1}_{d_k}$ are identity matrices of size $d_k\times d_k$ and $d_k$ is
the dimension of each one of the irreducible equivalent representations of 
$G_{\text{orb}}$ repeated in $V_{N,S,m_S}^{\chi_k}$.

\subsection{Proof of the assumption [Eq.~\eqref{assumsym}]}\label{appsym_sub}

Let us prove that Eq.~\eqref{assumsym} is verified 
for every group $G$ that does not mix 
configurations belonging to different eigenspaces of the number operator 
$\hat{N}$.
We need two preliminary observations. (i) By assumption our uncorrelated 
wavefunction $\ket{\Psi_0}$ is invariant respect to the action of $G$, 
i.e.,
\be
g\ket{\Psi_0}=e^{i\phi_g}\ket{\Psi_0}\quad\forall g\in G\,.
\ee
(ii) From Eq.~\eqref{trbands} we have that
\be
D^\dagger(g)\bar{\rho}^0 D(g)=\bar{\rho}^0\quad\forall g\in G\,,
\label{invdens}
\ee
where 
\be
\bar{\rho}^0_{\alpha\beta}\equiv\Av{\Psi_0}{\cc_\alpha\ca_\beta}
\ee
is the variational density matrix expressed in the original basis, 
see Eq.~\eqref{ldmcbasis}.
The density matrix $\bar{\rho}^0$
has exactly the same form of Eqs.~(\ref{formphiblock}-\ref{formphiblock3}) 
because of Eq.~\eqref{invdens}.  
For this reason it can be diagonalized by means of a matrix 
$\mathcal{U}$ of the same form, i.e.,
\be
D^\dagger(g)\mathcal{U}D(g)=\mathcal{U}\quad\forall g\in G\,.
\label{commsingleUG}
\ee
The single particle transformations induced by the matrices 
$\mathcal{U}$ and $D(g)$ into the Fock many body space evidently commute as a 
consequence of Eq.~\eqref{commsingleUG}. 
This concludes the proof of Eq.~\eqref{assumsym}.


\section{Self-consistent formulation of the $\phi$-matrix optimization}\label{philin}

We need to minimize the energy functional defined in 
Eq.~\eqref{variationalenergych1repeat} respect to the vector $c$ fulfilling the Gutzwiller 
constraints~[Eqs.~(\ref{uno-app-bisch1repeat}-\ref{due-app-bis1ch1repeat})].
The Gutzwiller constraints can be ensured by means of 
the following Lagrange functional
\be
\mathcal{L}[c,\lambda]=\sum_{\alpha\beta}\lambda_{\alpha\beta}
\Av{c}{N^S_{\alpha\beta}}\,.
\label{lagphiapp}
\ee

The variation of $\E_{\Psi_0}[c]$ is given by
\begin{widetext}
\bea
\delta\mathcal{E}[c]&=&
\Mel{\delta c}{\sum_{\alpha\beta}\frac{\partial\Av{\Psi_0}{\hat{T}^G[\Psi_0,c]}}
{\partial \R_{\alpha\beta}}\frac{{M}^S_{\alpha\beta}}{\sqrt{n^0_{\beta}(1-n^0_{\beta})}}}{c}
+\Mel{c}{\sum_{\alpha\beta}\frac{\partial\Av{\Psi_0}{{T}^G[\Psi_0,c]}}
{\partial \R_{\alpha\beta}}\frac{{M}^S_{\alpha\beta}}{\sqrt{n^0_{\beta}(1-n^0_{\beta})}}}{\delta c}
\nonumber\\
&+&\Mel{\delta c}{{U}}{c}+\Mel{c}{{U}}{\delta c}\,,
\eea
\end{widetext}
and the variation of the Lagrange functional~[Eq.~\eqref{lagphiapp}] is given by
\bea
\delta\mathcal{L}[c,\lambda]\!&=&\!
\Mel{\delta c}{\sum_{\alpha\beta}\lambda_{\alpha\beta}N^S_{\alpha\beta}}{c}\nonumber\\
\!&+&\!\Mel{c}{\sum_{\alpha\beta}\lambda_{\alpha\beta}N^S_{\alpha\beta}}{\delta c}\,.
\eea

The condition that the variation of 
\be
\delta\mathcal{F}_{\Psi_0}[c,\lambda]\equiv\delta\mathcal{E}_{\Psi_0}[c]
+\delta\mathcal{L}[c,\lambda]=0
\quad\forall\delta c\perp c
\ee
is equivalent to the following ``nonlinear eigenvalue problem''
\bea
{F}_{\Psi_0}[c,\lambda]\ket{c}=E\ket{c}\,,
\eea
where
\begin{widetext}
\bea
{F}_{\Psi_0}[c,\lambda]=U+
\sum_{\alpha\beta}\frac{\partial\Av{\Psi_0}{\hat{T}^G[\Psi_0,c]}}
{\partial \R_{\alpha\beta}}\frac{M^S_{\alpha\beta}}{\sqrt{n^0_{\beta}(1-n^0_{\beta})}}
+\sum_{\alpha\beta}\lambda_{\alpha\beta}N^S_{\alpha\beta}\,.
\eea
\end{widetext}

In principle, the minimization of $\E_{\Psi_0}[c]$ could be performed recursively,
starting from a given ``guess'' ${c_0}$ and iterating the following 
eigenvalue problem
\be
{F}_{\Psi_0}[c_n,\lambda_n]\ket{c_{n+1}}=E_{n+1}\ket{c_{n+1}}\,,
\label{fnnp1}
\ee
where $E_{n+1}$ is the lowest eigenvalue of ${F}_{\Psi_0}[c_n,\lambda_n]$
and $\lambda_n$ are the Lagrange multipliers such that ${c_{n+1}}$
satisfies the Gutzwiller constraints.
The minimum of $\E_{\Psi_0}[c]$ is realized in 
\be
c_{\text{min}}=\lim_{n\rightarrow\infty}c_n\,.
\ee

Instead to calculate $c_{\text{min}}$, it is convenient 
to approximate $c_{\text{min}}$ with ${c_1}$. 
This approximation can be considered as the result of a ``linearization''
of the functional $\E_{\Psi_0}[c]$ around the initial guess ${c_0}$.

\section{Numerical implementation of the tight binding problem}

Let us consider a general translational invariant non-interacting
tight binding Hamiltonian
\be
\hat{H}=\hat{T}+\Delta\hat{H}
\label{hamtightall}
\ee
where
\bea
\hat{T}&=&\sum_{\alpha\beta}\sum_{\bR\neq\bRp}\,t_{\bR\bRp}^{\alpha\beta}\,
\dc_{\bR\alpha}\da_{\bRp\beta}
\label{hamtightbindinggen}\\
\Delta\hat{H}&=&\sum_{\alpha\beta}\delta^{\alpha\beta}\sum_{\bR}\,
\dc_{\bR\alpha}\da_{\bR\beta}\,.
\label{hamlocalgen}
\eea
The translational invariance of the system
\be
t_{\bR\bRp}^{\alpha\beta}=t_{\bR+\bR_0\,\bRp+\bR_0}^{\alpha\beta}
\quad\forall\, \bR_0,\,\alpha,\beta 
\ee
allows to express $\hat{H}$ in $k$-space
\be
\hat{H}=\sum_{\alpha\beta}\sum_{\bk}
\left(t_{\bk}^{\alpha\beta}+\delta^{\alpha\beta}\right)
\dc_{\bk\alpha}\da_{\bk\beta}
\,,
\label{hamtightbindinggenkspace}
\ee
where
\be
t_{\bk}^{\alpha\beta}=\sum_\bR e^{-i\bk\bR}\,t_{\bR0}^{\alpha\beta}\,.
\ee
From Eq.~\eqref{hamtightbindinggenkspace} $\hat{H}$ can be easily diagonalized
numerically and expressed in terms of it's eigenoperators as follows
\be
\hat{H}=\sum_{\bk n}\epsilon_{\bk n}\,
\eta^\dagger_{\bk n}\eta^\dagga_{\bk n}\,.
\ee
Notice that the overlap matrix
\be
(U_\bk)_{\alpha n}=\Av{0}{\da_{\bk\alpha}\eta^\dagger_{\bk n}}
\ee
appears in Eq.~\eqref{trmateqs2}.

\bibliographystyle{apsrev}

\begin{thebibliography}{55}
\expandafter\ifx\csname natexlab\endcsname\relax\def\natexlab#1{#1}\fi
\expandafter\ifx\csname bibnamefont\endcsname\relax
  \def\bibnamefont#1{#1}\fi
\expandafter\ifx\csname bibfnamefont\endcsname\relax
  \def\bibfnamefont#1{#1}\fi
\expandafter\ifx\csname citenamefont\endcsname\relax
  \def\citenamefont#1{#1}\fi
\expandafter\ifx\csname url\endcsname\relax
  \def\url#1{\texttt{#1}}\fi
\expandafter\ifx\csname urlprefix\endcsname\relax\def\urlprefix{URL }\fi
\providecommand{\bibinfo}[2]{#2}
\providecommand{\eprint}[2][]{\url{#2}}

\bibitem[{\citenamefont{Gutzwiller}(1963)}]{Gutzwiller1}
\bibinfo{author}{\bibfnamefont{M.~C.} \bibnamefont{Gutzwiller}},
  \bibinfo{journal}{Phys. Rev. Lett.} \textbf{\bibinfo{volume}{10}},
  \bibinfo{pages}{159} (\bibinfo{year}{1963}).

\bibitem[{\citenamefont{Gutzwiller}(1964)}]{Gutzwiller2}
\bibinfo{author}{\bibfnamefont{M.~C.} \bibnamefont{Gutzwiller}},
  \bibinfo{journal}{Phys. Rev.} \textbf{\bibinfo{volume}{134}},
  \bibinfo{pages}{A923} (\bibinfo{year}{1964}).

\bibitem[{\citenamefont{Gutzwiller}(1965)}]{Gutzwiller3}
\bibinfo{author}{\bibfnamefont{M.~C.} \bibnamefont{Gutzwiller}},
  \bibinfo{journal}{Phys. Rev.} \textbf{\bibinfo{volume}{137}},
  \bibinfo{pages}{A1726} (\bibinfo{year}{1965}).

\bibitem[{\citenamefont{Yokoyama and Shiba}(1987)}]{Yokuyama}
\bibinfo{author}{\bibfnamefont{H.}~\bibnamefont{Yokoyama}} \bibnamefont{and}
  \bibinfo{author}{\bibfnamefont{H.}~\bibnamefont{Shiba}}, \bibinfo{journal}{J.
  Phys. Soc. Jpn.} \textbf{\bibinfo{volume}{56}}, \bibinfo{pages}{1490}
  (\bibinfo{year}{1987}).

\bibitem[{\citenamefont{Capello et~al.}(2005)\citenamefont{Capello, Becca,
  Fabrizio, Sorella, and Tosatti}}]{Manuela}
\bibinfo{author}{\bibfnamefont{M.}~\bibnamefont{Capello}},
  \bibinfo{author}{\bibfnamefont{F.}~\bibnamefont{Becca}},
  \bibinfo{author}{\bibfnamefont{M.}~\bibnamefont{Fabrizio}},
  \bibinfo{author}{\bibfnamefont{S.}~\bibnamefont{Sorella}}, \bibnamefont{and}
  \bibinfo{author}{\bibfnamefont{E.}~\bibnamefont{Tosatti}},
  \bibinfo{journal}{Phys. Rev. Lett.} \textbf{\bibinfo{volume}{94}},
  \bibinfo{pages}{026406} (\bibinfo{year}{2005}).

\bibitem[{\citenamefont{Georges et~al.}(1996)\citenamefont{Georges, Kotliar,
  Krauth, and Rozenberg}}]{DMFT}
\bibinfo{author}{\bibfnamefont{A.}~\bibnamefont{Georges}},
  \bibinfo{author}{\bibfnamefont{G.}~\bibnamefont{Kotliar}},
  \bibinfo{author}{\bibfnamefont{W.}~\bibnamefont{Krauth}}, \bibnamefont{and}
  \bibinfo{author}{\bibfnamefont{M.~J.} \bibnamefont{Rozenberg}},
  \bibinfo{journal}{Rev. Mod. Phys.} \textbf{\bibinfo{volume}{68}},
  \bibinfo{pages}{13} (\bibinfo{year}{1996}).

\bibitem[{\citenamefont{Metzner and Vollhardt}(1987)}]{Metzner-Vollhardt-PRL}
\bibinfo{author}{\bibfnamefont{W.}~\bibnamefont{Metzner}} \bibnamefont{and}
  \bibinfo{author}{\bibfnamefont{D.}~\bibnamefont{Vollhardt}},
  \bibinfo{journal}{Phys. Rev. Lett.} \textbf{\bibinfo{volume}{59}},
  \bibinfo{pages}{121} (\bibinfo{year}{1987}).

\bibitem[{\citenamefont{Metzner and Vollhardt}(1988)}]{Metzner-Vollhardt-PRB}
\bibinfo{author}{\bibfnamefont{W.}~\bibnamefont{Metzner}} \bibnamefont{and}
  \bibinfo{author}{\bibfnamefont{D.}~\bibnamefont{Vollhardt}},
  \bibinfo{journal}{Phys. Rev. B} \textbf{\bibinfo{volume}{37}},
  \bibinfo{pages}{7382} (\bibinfo{year}{1988}).

\bibitem[{\citenamefont{Gebhard}(1990)}]{Gebhard1}
\bibinfo{author}{\bibfnamefont{F.}~\bibnamefont{Gebhard}},
  \bibinfo{journal}{Phys. Rev. B} \textbf{\bibinfo{volume}{41}},
  \bibinfo{pages}{9452} (\bibinfo{year}{1990}).

\bibitem[{\citenamefont{M{\"{u}ller}-Hartmann}(1989)}]{Muller}
\bibinfo{author}{\bibfnamefont{E.}~\bibnamefont{M{\"{u}ller}-Hartmann}},
  \bibinfo{journal}{Z. Phys. B} \textbf{\bibinfo{volume}{76}},
  \bibinfo{pages}{211} (\bibinfo{year}{1989}).

\bibitem[{\citenamefont{Brinkman and Rice}(1970)}]{brinkman&rice}
\bibinfo{author}{\bibfnamefont{W.~F.} \bibnamefont{Brinkman}} \bibnamefont{and}
  \bibinfo{author}{\bibfnamefont{T.~M.} \bibnamefont{Rice}},
  \bibinfo{journal}{Phys. Rev. B} \textbf{\bibinfo{volume}{2}},
  \bibinfo{pages}{4302} (\bibinfo{year}{1970}).

\bibitem[{\citenamefont{Dzierzawa et~al.}(1997)\citenamefont{Dzierzawa,
  Baeriswyl, and Martelo}}]{Dzierzawa}
\bibinfo{author}{\bibfnamefont{M.}~\bibnamefont{Dzierzawa}},
  \bibinfo{author}{\bibfnamefont{D.}~\bibnamefont{Baeriswyl}},
  \bibnamefont{and} \bibinfo{author}{\bibfnamefont{S.~M.}
  \bibnamefont{Martelo}}, \bibinfo{journal}{Helv. Phys. Acta}
  \textbf{\bibinfo{volume}{70}}, \bibinfo{pages}{124} (\bibinfo{year}{1997}).

\bibitem[{\citenamefont{Lanat\`a}(2010)}]{metr}
\bibinfo{author}{\bibfnamefont{N.}~\bibnamefont{Lanat\`a}},
  \bibinfo{journal}{Phys. Rev. B} \textbf{\bibinfo{volume}{82}},
  \bibinfo{pages}{195326} (\bibinfo{year}{2010}).

\bibitem[{\citenamefont{Schir\`o and Fabrizio}(2010)}]{PhysRevLett.105.076401}
\bibinfo{author}{\bibfnamefont{M.}~\bibnamefont{Schir\`o}} \bibnamefont{and}
  \bibinfo{author}{\bibfnamefont{M.}~\bibnamefont{Fabrizio}},
  \bibinfo{journal}{Phys. Rev. Lett.} \textbf{\bibinfo{volume}{105}},
  \bibinfo{pages}{076401} (\bibinfo{year}{2010}).

\bibitem[{\citenamefont{Lanat\`a and Strand}(2011)}]{td-gutz-transport}
\bibinfo{author}{\bibfnamefont{N.}~\bibnamefont{Lanat\`a}} \bibnamefont{and}
  \bibinfo{author}{\bibfnamefont{U.~R.} \bibnamefont{Strand}}
  (\bibinfo{year}{2011}), \eprint{cond-mat/1102.2741}.

\bibitem[{\citenamefont{Hohenberg and Kohn}(1964)}]{HohenbergandKohn}
\bibinfo{author}{\bibfnamefont{P.}~\bibnamefont{Hohenberg}} \bibnamefont{and}
  \bibinfo{author}{\bibfnamefont{W.}~\bibnamefont{Kohn}},
  \bibinfo{journal}{Phys. Rev.} \textbf{\bibinfo{volume}{136}},
  \bibinfo{pages}{B864} (\bibinfo{year}{1964}).

\bibitem[{\citenamefont{Kohn and Sham}(1965)}]{KohnandSham}
\bibinfo{author}{\bibfnamefont{W.}~\bibnamefont{Kohn}} \bibnamefont{and}
  \bibinfo{author}{\bibfnamefont{L.~J.} \bibnamefont{Sham}},
  \bibinfo{journal}{Phys. Rev.} \textbf{\bibinfo{volume}{140}},
  \bibinfo{pages}{A1133} (\bibinfo{year}{1965}).

\bibitem[{\citenamefont{Gunnarsson and Lundqvist}(1976)}]{LDA}
\bibinfo{author}{\bibfnamefont{O.}~\bibnamefont{Gunnarsson}} \bibnamefont{and}
  \bibinfo{author}{\bibfnamefont{B.~I.} \bibnamefont{Lundqvist}},
  \bibinfo{journal}{Phys. Rev. B} \textbf{\bibinfo{volume}{13}},
  \bibinfo{pages}{4274} (\bibinfo{year}{1976}).

\bibitem[{\citenamefont{Ho et~al.}(2008)\citenamefont{Ho, Schmalian, and
  Wang}}]{poorchguy}
\bibinfo{author}{\bibfnamefont{K.~M.} \bibnamefont{Ho}},
  \bibinfo{author}{\bibfnamefont{J.}~\bibnamefont{Schmalian}},
  \bibnamefont{and} \bibinfo{author}{\bibfnamefont{C.~Z.} \bibnamefont{Wang}},
  \bibinfo{journal}{Phys. Rev. B} \textbf{\bibinfo{volume}{77}},
  \bibinfo{pages}{073101} (\bibinfo{year}{2008}).

\bibitem[{\citenamefont{Deng et~al.}(2009)\citenamefont{Deng, Wang, Dai, and
  Fang}}]{Fang}
\bibinfo{author}{\bibfnamefont{X.~Y.} \bibnamefont{Deng}},
  \bibinfo{author}{\bibfnamefont{L.}~\bibnamefont{Wang}},
  \bibinfo{author}{\bibfnamefont{X.}~\bibnamefont{Dai}}, \bibnamefont{and}
  \bibinfo{author}{\bibfnamefont{Z.}~\bibnamefont{Fang}},
  \bibinfo{journal}{Phys. Rev. B} \textbf{\bibinfo{volume}{79}},
  \bibinfo{eid}{075114} (\bibinfo{year}{2009}).

\bibitem[{\citenamefont{Anisimov et~al.}(1997)\citenamefont{Anisimov,
  Aryasetiawan, and Lichtenstein}}]{LDA+U}
\bibinfo{author}{\bibfnamefont{V.~I.} \bibnamefont{Anisimov}},
  \bibinfo{author}{\bibfnamefont{F.}~\bibnamefont{Aryasetiawan}},
  \bibnamefont{and}
  \bibinfo{author}{\bibfnamefont{A.}~\bibnamefont{Lichtenstein}},
  \bibinfo{journal}{J. Phys. Condens. Matter} \textbf{\bibinfo{volume}{9}},
  \bibinfo{pages}{767} (\bibinfo{year}{1997}).

\bibitem[{\citenamefont{Kotliar et~al.}(2006)\citenamefont{Kotliar, Savrasov,
  Haule, Oudovenko, Parcollet, and Marianetti}}]{LDA+U+DMFT}
\bibinfo{author}{\bibfnamefont{G.}~\bibnamefont{Kotliar}},
  \bibinfo{author}{\bibfnamefont{S.~Y.} \bibnamefont{Savrasov}},
  \bibinfo{author}{\bibfnamefont{K.}~\bibnamefont{Haule}},
  \bibinfo{author}{\bibfnamefont{V.~S.} \bibnamefont{Oudovenko}},
  \bibinfo{author}{\bibfnamefont{O.}~\bibnamefont{Parcollet}},
  \bibnamefont{and} \bibinfo{author}{\bibfnamefont{C.~A.}
  \bibnamefont{Marianetti}}, \bibinfo{journal}{Rev. Mod. Phys.}
  \textbf{\bibinfo{volume}{78}}, \bibinfo{eid}{865} (\bibinfo{year}{2006}).

\bibitem[{\citenamefont{B\"unemann et~al.}(1998)\citenamefont{B\"unemann,
  Weber, and Gebhard}}]{Gebhard}
\bibinfo{author}{\bibfnamefont{J.}~\bibnamefont{B\"unemann}},
  \bibinfo{author}{\bibfnamefont{W.}~\bibnamefont{Weber}}, \bibnamefont{and}
  \bibinfo{author}{\bibfnamefont{F.}~\bibnamefont{Gebhard}},
  \bibinfo{journal}{Phys. Rev. B} \textbf{\bibinfo{volume}{57}},
  \bibinfo{pages}{6896} (\bibinfo{year}{1998}).

\bibitem[{\citenamefont{Kanamori}(1963)}]{Kanamori}
\bibinfo{author}{\bibfnamefont{J.}~\bibnamefont{Kanamori}},
  \bibinfo{journal}{Proj. Theor. Phys.} \textbf{\bibinfo{volume}{30}},
  \bibinfo{pages}{275} (\bibinfo{year}{1963}).

\bibitem[{\citenamefont{Werner et~al.}(2009)\citenamefont{Werner, Gull, and
  Millis}}]{Werner-PRB2009}
\bibinfo{author}{\bibfnamefont{P.}~\bibnamefont{Werner}},
  \bibinfo{author}{\bibfnamefont{E.}~\bibnamefont{Gull}}, \bibnamefont{and}
  \bibinfo{author}{\bibfnamefont{A.~J.} \bibnamefont{Millis}},
  \bibinfo{journal}{Phys. Rev. B} \textbf{\bibinfo{volume}{79}},
  \bibinfo{pages}{115119} (\bibinfo{year}{2009}).

\bibitem[{\citenamefont{Costi and Liebsch}(2007)}]{Costi-PRL2007}
\bibinfo{author}{\bibfnamefont{T.~A.} \bibnamefont{Costi}} \bibnamefont{and}
  \bibinfo{author}{\bibfnamefont{A.}~\bibnamefont{Liebsch}},
  \bibinfo{journal}{Phys. Rev. Lett.} \textbf{\bibinfo{volume}{99}},
  \bibinfo{pages}{236404} (\bibinfo{year}{2007}).

\bibitem[{\citenamefont{Werner et~al.}(2008)\citenamefont{Werner, Gull, Troyer,
  and Millis}}]{Werner2008aa}
\bibinfo{author}{\bibfnamefont{P.}~\bibnamefont{Werner}},
  \bibinfo{author}{\bibfnamefont{E.}~\bibnamefont{Gull}},
  \bibinfo{author}{\bibfnamefont{M.}~\bibnamefont{Troyer}}, \bibnamefont{and}
  \bibinfo{author}{\bibfnamefont{A.~J.} \bibnamefont{Millis}},
  \bibinfo{journal}{Phys. Rev. Lett.} \textbf{\bibinfo{volume}{101}},
  \bibinfo{pages}{166405} (\bibinfo{year}{2008}).

\bibitem[{\citenamefont{B\"unemann and Weber}(1997)}]{PhysRevB.55.4011}
\bibinfo{author}{\bibfnamefont{J.}~\bibnamefont{B\"unemann}} \bibnamefont{and}
  \bibinfo{author}{\bibfnamefont{W.}~\bibnamefont{Weber}},
  \bibinfo{journal}{Phys. Rev. B} \textbf{\bibinfo{volume}{55}},
  \bibinfo{pages}{4011} (\bibinfo{year}{1997}).

\bibitem[{\citenamefont{Gul\'acsi et~al.}(1993)\citenamefont{Gul\'acsi, Strack,
  and Vollhardt}}]{PhysRevB.47.8594}
\bibinfo{author}{\bibfnamefont{Z.}~\bibnamefont{Gul\'acsi}},
  \bibinfo{author}{\bibfnamefont{R.}~\bibnamefont{Strack}}, \bibnamefont{and}
  \bibinfo{author}{\bibfnamefont{D.}~\bibnamefont{Vollhardt}},
  \bibinfo{journal}{Phys. Rev. B} \textbf{\bibinfo{volume}{47}},
  \bibinfo{pages}{8594} (\bibinfo{year}{1993}).

\bibitem[{\citenamefont{Attaccalite and Fabrizio}(2003)}]{Attaccalite}
\bibinfo{author}{\bibfnamefont{C.}~\bibnamefont{Attaccalite}} \bibnamefont{and}
  \bibinfo{author}{\bibfnamefont{M.}~\bibnamefont{Fabrizio}},
  \bibinfo{journal}{Phys. Rev. B} \textbf{\bibinfo{volume}{68}},
  \bibinfo{pages}{155117} (\bibinfo{year}{2003}).

\bibitem[{\citenamefont{Wang et~al.}(2006)\citenamefont{Wang, Wang, Chen, and
  Zhang}}]{wang:2006}
\bibinfo{author}{\bibfnamefont{Q.-H.} \bibnamefont{Wang}},
  \bibinfo{author}{\bibfnamefont{Z.~D.} \bibnamefont{Wang}},
  \bibinfo{author}{\bibfnamefont{Y.}~\bibnamefont{Chen}}, \bibnamefont{and}
  \bibinfo{author}{\bibfnamefont{F.~C.} \bibnamefont{Zhang}},
  \bibinfo{journal}{Phys. Rev. B} \textbf{\bibinfo{volume}{73}},
  \bibinfo{eid}{092507} (\bibinfo{year}{2006}).

\bibitem[{\citenamefont{Fabrizio}(2007)}]{fab}
\bibinfo{author}{\bibfnamefont{M.}~\bibnamefont{Fabrizio}},
  \bibinfo{journal}{Phys. Rev. B} \textbf{\bibinfo{volume}{76}},
  \bibinfo{eid}{165110} (\bibinfo{year}{2007}).

\bibitem[{\citenamefont{Lanat\`{a} et~al.}(2008)\citenamefont{Lanat\`{a},
  Barone, and Fabrizio}}]{lanata}
\bibinfo{author}{\bibfnamefont{N.}~\bibnamefont{Lanat\`{a}}},
  \bibinfo{author}{\bibfnamefont{P.}~\bibnamefont{Barone}}, \bibnamefont{and}
  \bibinfo{author}{\bibfnamefont{M.}~\bibnamefont{Fabrizio}},
  \bibinfo{journal}{Phys. Rev. B} \textbf{\bibinfo{volume}{78}},
  \bibinfo{eid}{155127} (\bibinfo{year}{2008}).

\bibitem[{\citenamefont{Lanat\`a et~al.}(2009)\citenamefont{Lanat\`a, Barone,
  and Fabrizio}}]{mybil}
\bibinfo{author}{\bibfnamefont{N.}~\bibnamefont{Lanat\`a}},
  \bibinfo{author}{\bibfnamefont{P.}~\bibnamefont{Barone}}, \bibnamefont{and}
  \bibinfo{author}{\bibfnamefont{M.}~\bibnamefont{Fabrizio}},
  \bibinfo{journal}{Phys. Rev. B} \textbf{\bibinfo{volume}{80}},
  \bibinfo{pages}{224524} (\bibinfo{year}{2009}).

\bibitem[{\citenamefont{Ferrero et~al.}(2005)\citenamefont{Ferrero, Becca,
  Fabrizio, and Capone}}]{ferreroPRB}
\bibinfo{author}{\bibfnamefont{M.}~\bibnamefont{Ferrero}},
  \bibinfo{author}{\bibfnamefont{F.}~\bibnamefont{Becca}},
  \bibinfo{author}{\bibfnamefont{M.}~\bibnamefont{Fabrizio}}, \bibnamefont{and}
  \bibinfo{author}{\bibfnamefont{M.}~\bibnamefont{Capone}},
  \bibinfo{journal}{Phys. Rev. B} \textbf{\bibinfo{volume}{72}},
  \bibinfo{eid}{205126} (\bibinfo{year}{2005}).

\bibitem[{\citenamefont{Lanat\`a}(2009)}]{mythesis}
\bibinfo{author}{\bibfnamefont{N.}~\bibnamefont{Lanat\`a}}, Ph.D. thesis,
  \bibinfo{school}{SISSA-Trieste} (\bibinfo{year}{2009}).

\bibitem[{\citenamefont{Wigner}(1959)}]{Wigner}
\bibinfo{author}{\bibfnamefont{E.~P.} \bibnamefont{Wigner}},
  \emph{\bibinfo{title}{Group theory and its application to the quantum
  mechanics of atomic spectra}} (\bibinfo{publisher}{Academic Press},
  \bibinfo{year}{1959}).

\bibitem[{\citenamefont{Levy}(1979)}]{324}
\bibinfo{author}{\bibfnamefont{M.}~\bibnamefont{Levy}}, \bibinfo{journal}{Proc.
  Nat. Acad. Sci.} \textbf{\bibinfo{volume}{76}}, \bibinfo{pages}{6062}
  (\bibinfo{year}{1979}).

\bibitem[{\citenamefont{Levy}(1982)}]{325}
\bibinfo{author}{\bibfnamefont{M.}~\bibnamefont{Levy}}, \bibinfo{journal}{Phys.
  Rev. A} \textbf{\bibinfo{volume}{26}}, \bibinfo{pages}{1200}
  (\bibinfo{year}{1982}).

\bibitem[{\citenamefont{Lieb}(1983)}]{327}
\bibinfo{author}{\bibfnamefont{E.}~\bibnamefont{Lieb}}, \bibinfo{journal}{Int.
  J. Quant. Chem.} \textbf{\bibinfo{volume}{24}}, \bibinfo{pages}{243}
  (\bibinfo{year}{1983}).

\bibitem[{\citenamefont{Heath}(2002)}]{mtheath}
\bibinfo{author}{\bibfnamefont{M.~T.} \bibnamefont{Heath}},
  \emph{\bibinfo{title}{Scientific Computing An Introductory Survey}}
  (\bibinfo{publisher}{McGraw-Hill Book Company}, \bibinfo{year}{2002}),
  \bibinfo{edition}{2nd} ed.

\bibitem[{\citenamefont{Borghi et~al.}(2010)\citenamefont{Borghi, Fabrizio, and
  Tosatti}}]{borghi}
\bibinfo{author}{\bibfnamefont{G.}~\bibnamefont{Borghi}},
  \bibinfo{author}{\bibfnamefont{M.}~\bibnamefont{Fabrizio}}, \bibnamefont{and}
  \bibinfo{author}{\bibfnamefont{E.}~\bibnamefont{Tosatti}},
  \bibinfo{journal}{Phys. Rev. B} \textbf{\bibinfo{volume}{81}},
  \bibinfo{pages}{115134} (\bibinfo{year}{2010}).

\bibitem[{\citenamefont{Lechermann et~al.}(2007)\citenamefont{Lechermann,
  Georges, Kotliar, and Parcollet}}]{Lechermann:2007ys}
\bibinfo{author}{\bibfnamefont{F.}~\bibnamefont{Lechermann}},
  \bibinfo{author}{\bibfnamefont{A.}~\bibnamefont{Georges}},
  \bibinfo{author}{\bibfnamefont{G.}~\bibnamefont{Kotliar}}, \bibnamefont{and}
  \bibinfo{author}{\bibfnamefont{O.}~\bibnamefont{Parcollet}},
  \bibinfo{journal}{Phys. Rev. B} \textbf{\bibinfo{volume}{76}},
  \bibinfo{pages}{155102} (\bibinfo{year}{2007}).

\bibitem[{\citenamefont{Barnes}(1976)}]{Barnes1}
\bibinfo{author}{\bibfnamefont{S.~E.} \bibnamefont{Barnes}},
  \bibinfo{journal}{J. Phys. F: Met. Phys.} \textbf{\bibinfo{volume}{6}},
  \bibinfo{pages}{1375} (\bibinfo{year}{1976}).

\bibitem[{\citenamefont{Barnes}(1977)}]{Barnes2}
\bibinfo{author}{\bibfnamefont{S.~E.} \bibnamefont{Barnes}},
  \bibinfo{journal}{J. Phys. F: Met. Phys.} \textbf{\bibinfo{volume}{7}},
  \bibinfo{pages}{2637} (\bibinfo{year}{1977}).

\bibitem[{\citenamefont{Coleman}(1983)}]{Coleman-1/N}
\bibinfo{author}{\bibfnamefont{P.}~\bibnamefont{Coleman}},
  \bibinfo{journal}{Phys. Rev. B} \textbf{\bibinfo{volume}{28}},
  \bibinfo{pages}{5255} (\bibinfo{year}{1983}).

\bibitem[{\citenamefont{Read and Newns}(1983)}]{Read&Newns}
\bibinfo{author}{\bibfnamefont{N.}~\bibnamefont{Read}} \bibnamefont{and}
  \bibinfo{author}{\bibfnamefont{D.~M.} \bibnamefont{Newns}},
  \bibinfo{journal}{J Phys. C: Solid State Phys.}
  \textbf{\bibinfo{volume}{16}}, \bibinfo{pages}{3273} (\bibinfo{year}{1983}).

\bibitem[{\citenamefont{Kotliar and Ruckenstein}(1986)}]{Kotliar-Ruckenstein}
\bibinfo{author}{\bibfnamefont{G.}~\bibnamefont{Kotliar}} \bibnamefont{and}
  \bibinfo{author}{\bibfnamefont{A.~E.} \bibnamefont{Ruckenstein}},
  \bibinfo{journal}{Phys. Rev. Lett.} \textbf{\bibinfo{volume}{57}},
  \bibinfo{pages}{1362} (\bibinfo{year}{1986}).

\bibitem[{\citenamefont{Bunemann and Gebhard}(2007)}]{quaquaraqua}
\bibinfo{author}{\bibfnamefont{J.}~\bibnamefont{Bunemann}} \bibnamefont{and}
  \bibinfo{author}{\bibfnamefont{F.}~\bibnamefont{Gebhard}},
  \bibinfo{journal}{Phys. Rev.B} \textbf{\bibinfo{volume}{76}},
  \bibinfo{eid}{193104} (\bibinfo{year}{2007}).

\bibitem[{\citenamefont{B\"unemann et~al.}(2003)\citenamefont{B\"unemann,
  Gebhard, and Thul}}]{Gebhard-FL}
\bibinfo{author}{\bibfnamefont{J.}~\bibnamefont{B\"unemann}},
  \bibinfo{author}{\bibfnamefont{F.}~\bibnamefont{Gebhard}}, \bibnamefont{and}
  \bibinfo{author}{\bibfnamefont{R.}~\bibnamefont{Thul}},
  \bibinfo{journal}{Phys. Rev. B} \textbf{\bibinfo{volume}{67}},
  \bibinfo{pages}{075103} (\bibinfo{year}{2003}).

\bibitem[{\citenamefont{Koga et~al.}(2002)\citenamefont{Koga, Imai, and
  Kawakami}}]{PhysRevB.66.165107}
\bibinfo{author}{\bibfnamefont{A.}~\bibnamefont{Koga}},
  \bibinfo{author}{\bibfnamefont{Y.}~\bibnamefont{Imai}}, \bibnamefont{and}
  \bibinfo{author}{\bibfnamefont{N.}~\bibnamefont{Kawakami}},
  \bibinfo{journal}{Phys. Rev. B} \textbf{\bibinfo{volume}{66}},
  \bibinfo{pages}{165107} (\bibinfo{year}{2002}).

\bibitem[{\citenamefont{Werner and Millis}(2007)}]{Werner:2007lr}
\bibinfo{author}{\bibfnamefont{P.}~\bibnamefont{Werner}} \bibnamefont{and}
  \bibinfo{author}{\bibfnamefont{A.~J.} \bibnamefont{Millis}},
  \bibinfo{journal}{Phys. Rev. Lett.} \textbf{\bibinfo{volume}{99}},
  \bibinfo{pages}{126405} (\bibinfo{year}{2007}).

\bibitem[{\citenamefont{L\"auchli and Werner}(2009)}]{Lauchli:2009fk}
\bibinfo{author}{\bibfnamefont{A.~M.} \bibnamefont{L\"auchli}}
  \bibnamefont{and} \bibinfo{author}{\bibfnamefont{P.}~\bibnamefont{Werner}},
  \bibinfo{journal}{Phys. Rev. B} \textbf{\bibinfo{volume}{80}},
  \bibinfo{pages}{235117} (\bibinfo{year}{2009}).

\bibitem[{\citenamefont{Strand et~al.}(2011)\citenamefont{Strand, Sabashvili,
  Granath, Hellsing, and \"Ostlund}}]{hugo}
\bibinfo{author}{\bibfnamefont{H.~U.~R.} \bibnamefont{Strand}},
  \bibinfo{author}{\bibfnamefont{A.}~\bibnamefont{Sabashvili}},
  \bibinfo{author}{\bibfnamefont{M.}~\bibnamefont{Granath}},
  \bibinfo{author}{\bibfnamefont{B.}~\bibnamefont{Hellsing}}, \bibnamefont{and}
  \bibinfo{author}{\bibfnamefont{S.}~\bibnamefont{\"Ostlund}},
  \bibinfo{journal}{Phys. Rev. B} \textbf{\bibinfo{volume}{83}},
  \bibinfo{pages}{205136} (\bibinfo{year}{2011}).

\bibitem[{\citenamefont{Zitko}(2009)}]{zitko}
\bibinfo{author}{\bibfnamefont{R.}~\bibnamefont{Zitko}},
  \bibinfo{journal}{Phys. Rev. B} \textbf{\bibinfo{volume}{80}},
  \bibinfo{pages}{125125} (\bibinfo{year}{2009}).

\end{thebibliography}

\end{document}